%% file: main.tex
\documentclass[conference]{IEEEtran}
\IEEEoverridecommandlockouts
\usepackage{amsmath,amsfonts}
\usepackage{algorithmic}
\usepackage{graphicx}
\usepackage{textcomp}
\usepackage{xcolor}
\usepackage{booktabs}
\usepackage{xcolor}
\usepackage{pgfplots}
\usepackage{subcaption}
\usepackage{colortbl}
\usepackage{braket}
\usepackage{epsfig,latexsym,graphics}
\usepackage[font={small}]{caption}
\usepackage{setspace}
\usepackage{enumitem}
\pgfplotsset{compat=1.16}
\usepackage{cmd}
\usepackage[spacesep,definevectors]{easyvector} 

\usepackage{comment}
\usepackage{balance}

\def\BibTeX{{\rm B\kern-.05em{\sc i\kern-.025em b}\kern-.08em
    T\kern-.1667em\lower.7ex\hbox{E}\kern-.125emX}}
    
\begin{document}

\title{Effective and General Distance Computation for Approximate Nearest Neighbor Search}
%Data-Driven Distance Computation for Approximate Nearest Neighbor Search

\author{\IEEEauthorblockN{Mingyu Yang$^{1,3}$, Wentao Li$^{1,4}$, Jiabao Jin$^{2}$, Xiaoyao Zhong$^{2}$, Xiangyu Wang$^{2}$,\\ Zhitao Shen$^{2}$, Wei Jia$^{2}$,  and Wei Wang$^{1,3}$}
\IEEEauthorblockA{$^1$The Hong Kong University of Science and Technology (Guangzhou); $^2$Ant Group; \\$^3$The Hong Kong University of Science and Technology; $^4$University of Leicester\\
    myang250@connect.hkust-gz.edu.cn;
    wentaoli@hkust-gz.edu.au; \\
    \{jinjiabao.jjb, zhongxiaoyao.zxy, wxy407827, zhitao.szt, jw94525\}@antgroup.com; 
    weiwcs@ust.hk
    }
}

\maketitle

\input{abstract}


\input{intro}

\input{preliminary}

\input{Methodology}

\input{analysis}

\input{exp}
\input{conclusion}

\input{acknowledgment}

\newpage
\balance
\bibliographystyle{IEEEtran}
\bibliography{reference}
\vspace{12pt}

\end{document}

%% file: abstract.tex
\begin{abstract}
Approximate K Nearest Neighbor (AKNN) search in high-dimensional spaces is a critical yet challenging problem. In AKNN search, distance computation is the core task that dominates the runtime. Existing approaches typically use approximate distances to improve computational efficiency, often at the cost of reduced search accuracy. To address this issue, the state-of-the-art method, $\ADS$, employs random projections to estimate approximate distances and introduces an additional distance correction process to mitigate accuracy loss. However, $\ADS$ has limitations in both \textit{effectiveness} and \textit{generality}, primarily due to its heavy reliance on random projections for distance approximation and correction.

Motivated by this, we leverage data distribution to improve distance approximation via orthogonal projection, thereby addressing the effectiveness limitation of $\ADS$;
we also adopt a data-driven approach to distance correction, decoupling the correction process from the distance approximation process, thereby overcoming the generality limitation of $\ADS$.
Extensive experiments demonstrate the superiority and effectiveness of our method. 
In particular, compared to $\ADS$, our method achieves a speedup of 1.6 to 2.1 times on real-world datasets while providing higher accuracy. In addition, our method shows superior performance in Ant Group image search scenarios and has been integrated into their search engine.
%To address the effectiveness limitation of $\ADS$, we leverage data distribution to improve distance approximation via orthogonal projection. 
%Furthermore, to overcome the generality limitation of $\ADS$, we adopt a data-driven approach to distance correction, decoupling the correction process from the distance approximation process. 
%Extensive experiments demonstrate the superiority and effectiveness of our method. 
%In particular, compared to $\ADS$, our method achieves a speedup of 1.6 to 2.1 times on real-world datasets while providing higher accuracy. In addition, our method shows superior performance in security search scenarios of Ant Group and has been integrated into their search engine.
\end{abstract}

%% file: intro.tex
\section{Introduction}
The problem of the K Nearest Neighbor (KNN) search involves identifying the top-K data points in a database $S$ that are closest to a query point $q$. 
KNN search is crucial in various domains, including information retrieval~\cite{Image-Search-2007-PR}, data mining~\cite{NN-datamining-1967-TIT}, recommender systems~\cite{CF-2007-recommender-sys}, and vector databases~\cite{Vector-database-Liguoliang-2024-SIGMOD}.
Effective solutions, such as R-trees, exist for KNN search in low-dimensional spaces. 
However, the curse of dimensionality~\cite{Curse-of-dim-1998} renders exact KNN search prohibitively time-consuming in high-dimensional spaces. 
Consequently, researchers have developed the approximate variant known as \textbf{Approximate K Nearest Neighbors (AKNN) search}~\cite{Curse-of-dim-1998}, which is more suitable for real-time responses in large-scale data.

Given the critical role of AKNN search, numerous algorithms have been developed. These algorithms mainly fall into four categories: inverted file-based~\cite{DBLP:journals/pami/PQJegouDS11,IMI:DBLP:journals/pami/BabenkoL15}, graph-based~\cite{NSWDBLP:journals/is/MalkovPLK14,DBLP:journals/pami/HNSWMalkovY20,DBLP:journals/pami/SSGFuWC22,DBLP:journals/pvldb/NSGFuXWC19,DPG:DBLP:journals/tkde/LiZSWLZL20,tMRNG:journals/pacmmod/PengCCYX23}, tree-based~\cite{Random-Project-Tree-2008,Cover-Tree-2006-ICML,Revisit-KD-tree-2019-KDD}, and hash-based~\cite{SRS-yifang-2014,PMLSH-bolong-2020,C2LSH-2012-SIGMOD,RPLSH-1995,QALSH-2015-VLDB,query-aware-LSH} methods. 
To find the AKNN of a query point $q$ in a database $S$, these AKNN algorithms can often be abstracted into a \textbf{candidate generation and refinement framework}:
(1) Candidate generation: In this phase, a subset of points from $S$ is selected to form a superset of the final AKNN. 
(2) Refinement: In this phase, the algorithm identifies the top points closest to $q$ among the candidates, which are then returned as the AKNN.

The distinction between various AKNN algorithms primarily lies in the candidate generation phase, while the refinement phase is almost identical.
% For instance, inverted file-based methods, such as $\IVF$~\cite{IMI:DBLP:journals/pami/BabenkoL15}, utilize clustering, while graph-based methods, like $\HNSW$~\cite{DBLP:journals/pami/HNSWMalkovY20}, employ greedy traversal to generate candidates. 
% On the other hand, the refinement phase of AKNN algorithms is generally consistent across different approaches. 
In the refinement phase, a result queue $Q$, often implemented as a max-heap, is maintained to store the data points closest to the query point $q$, ultimately producing the final result. 
Specifically, for a candidate point $p$, if the distance to the query point $q$ is less than the maximum distance $\tau$ recorded in $Q$, the result queue is updated; otherwise, the point is disregarded. 
Therefore, distance computation is crucial and computationally intensive during this phase.
Notably, \textbf{distance computation is often the most time-consuming component of AKNN algorithms.}
For example, in graph-based algorithms like $\HNSW$~\cite{DBLP:journals/pami/HNSWMalkovY20}, distance computation constitutes 80\% of the total AKNN search time. 
In inverted file-based algorithms such as $\IVF$~\cite{IMI:DBLP:journals/pami/BabenkoL15}, it accounts for 90\% of the total time cost~\cite{ADSampling:journals/sigmod/GaoL23}. 
Thus, accelerating distance computation is essential for expediting AKNN search.

\stitle{The State-of-the-art.} 
To accelerate distance computation, $\ADS$~\cite{ADSampling:journals/sigmod/GaoL23} was introduced as a general plug-in. 
$\ADS$ first estimates an (initial) approximate distance between two points by random projection and obtains an error bound based on the random projection matrix used. 
The advantage of $\ADS$ lies in its ability to use error bounds for an additional distance correction process: it corrects the approximate distance by error bounds and analyzes whether the use of corrected approximate distances is sufficient in AKNN search~\cite{Learning-Hash-survey-2017-PAMI,ADSampling:journals/sigmod/GaoL23}. 
If not, more accurate distances are calculated, and an incremental correction is applied to the approximation (until an exact distance is computed).
$\ADS$ achieves a good balance between speed and accuracy in AKNN search by the correction process, and experimental results confirm its efficiency. 
Yet, there is still much room for improvement in its effectiveness and generality.

\sstitle{Effectiveness.}
$\ADS$ employs a projection method to approximate distances. 
Specifically, $\ADS$ utilizes a \textbf{random} projection matrix to compute these approximate distances. However, within projection methods, \textbf{random projection typically incurs a larger error than optimal orthogonal projection} between approximate and exact distances. 
It is important to note that if the approximate distance is sufficiently accurate, $\ADS$ can avoid the need for the more time-consuming incremental distance correction process. Therefore, a more accurate approximate distance is crucial.

\sstitle{Generality.}
The error bound is critical for $\ADS$, as it informs whether the current (corrected) approximate distances are sufficient for the refinement phase of the AKNN search.
%necessitate additional dimensionality sampling. 
Yet, deriving the error bound for $\ADS$ is determined by the random projection matrix employed in the distance estimation process. 
\textbf{This dependence constrains the applicability of $\ADS$} to scenarios where distances are approximated by techniques other than random projection.

\stitle{Our Idea.}
The limitation of $\ADS$ in both effectiveness and generality stems from its over-reliance on random projections. 
This dependency limits its potential as it overlooks the underlying properties of the data points in the database $S$ --- $\ADS$ relied only on the projection matrix, not the database itself, for distance correction.
This raises a natural question: Can we develop a new method for distance computation that not only provides the error bounds for distance correction to maintain accuracy but also optimizes efficiency by exploiting the properties of data points in the database $S$?

In this paper, we provide an affirmative answer to this problem. 
First, we investigate the cause of the limited effectiveness of $\ADS$. 
By decomposing the exact distance into approximate distance and estimation error, we prove that the optimal orthogonal projection, when applied to database points, minimizes the estimation error. 
This indicates that the random projection used by $\ADS$ is suboptimal.
We also analyze the distribution of estimation errors, and by introducing reasonable assumptions, we derive new error bounds for distance correction, which are shown to be minimized.

Furthermore, we introduce a novel data-driven distance correction scheme to accommodate the approximate distances generated by \textit{arbitrary} distance approximation methods. 
Our method is unique in that it makes no assumptions about the source of these approximate distances.
Instead, it parameterizes the error bound used in the distance correction process and learns this parameter directly from the data, adopting a data-driven approach. 
This flexibility allows our method to adapt to any approximate distance methods, such as those obtained from product quantization (PQ)~\cite{DBLP:journals/pami/PQJegouDS11}, thereby providing a level of generality that $\ADS$ lacks. 

\stitle{Contributions.}
We summarize our contributions as follows:

\sstitle{Analysis of the SOTA Method (\S~\ref{sec:problem}).}
We introduce the state-of-the-art distance computation method, $\ADS$, and analyze its limitations.
This analysis motivates us to propose more effective and general methods for distance computation.

\sstitle{More Effective Distance Estimation (\S~\ref{sec:estimation}).}
To address the limited effectiveness of $\ADS$, we decompose the exact distance into the approximate distance and the estimation error. 
We demonstrate that using optimal orthogonal (i.e., PCA) projections instead of random projections results in minimal estimation error. 
Consequently, we replace the random projections used in $\ADS$ with PCA projections to improve the accuracy of distance estimation. 
In addition, by analyzing the distribution of estimation errors and assuming a Gaussian distribution, we derive the corresponding optimal error bounds for distance correction.

\sstitle{More General Distance Correction (\S~\ref{sec:correction}).}
To accommodate non-projection-based distance estimation methods (such as PQ~\cite{DBLP:journals/pami/PQJegouDS11}), we propose a data-driven distance correction scheme. 
This scheme parameterizes the error bound used for distance correction and employs a data-driven approach to learn the appropriate parameter for a given database.
By determining this parameter/error bound using the learning-based technique, the proposed distance correction scheme avoids making assumptions about the source of the approximate distance, thus achieving the generality that $\ADS$ lacks.
%We also discuss how the proposed distance correction scheme can be integrated into existing AKNN algorithms to enhance their efficiency.

\sstitle{Extensive Experimental Analysis (\S~\ref{sec:experiment}).}
We have conducted extensive experiments on a large number of real-world datasets, ranging from $1$ million to $100$ million entries, to validate our method.
The experimental results demonstrate that the proposed method improves the search efficiency of existing AKNN algorithms by $1.6$ to $2.1$ times, significantly outperforming $\ADS$. 
Furthermore, our algorithm is scalable to larger datasets, further confirming its efficiency.

\stitle{Related Work.}
In the area of speeding up distance computations between points or vectors, two prominent methods are $\ADS$~\cite{ADSampling:journals/sigmod/GaoL23} and FINGER~\cite{Finger-WWW-2023}. 
$\ADS$ uses the Johnson-Lindenstrauss (JL) lemma~\cite{JL-extension-1986} to establish a probabilistic bound between approximate and exact distances to speed up distance computations. 
This method, based on a solid mathematical foundation, is widely applicable to various AKNN algorithms.
On the other hand, FINGER~\cite{Finger-WWW-2023} is specifically designed for graph-based algorithms such as $\HNSW$. 
While FINGER shows empirical success in improving search speed over $\HNSW$~\cite{DBLP:journals/pami/HNSWMalkovY20}, it comes with increased index construction time and higher space requirements. 
Thus, in this paper, we focus primarily on improving the performance of $\ADS$, while providing a comparative analysis of FINGER in \S~\ref{sec:experiment}.

Data-driven approaches have greatly advanced the field of AKNN search. 
In particular, recent work~\cite{learning-to-roate2019,reinforce-routing2023} uses learning-based techniques to predict the next node during graph traversal, allowing for more efficient navigation of the search space. 
Other methods~\cite{Steiner-Hardess-zeyu-wang-VLDB-2024,Improve-Early-Terminate-SIGMD-2024} employ learning-based strategies to estimate the difficulty of the AKNN search, allowing early stops to improve the efficiency.
In contrast, our work focuses on optimizing distance computations, making it compatible with various AKNN algorithms. 
By addressing the bottlenecks associated with distance computation, our work provides a comprehensive solution to improve the efficiency of the entire AKNN search process.

Due to space limitations, some proofs and experiments have been omitted and can be found in the technical report~\cite{technicalreport}.

%% file: preliminary.tex
\section{Preliminaries}
\S~\ref{sub:anns} presents the AKNN search problem and its associated AKNN algorithms. 
Then, \S~\ref{sub:distance} discusses the issue of distance computation, an essential component of AKNN search.

\subsection{The AKNN Search}\label{sub:anns}
Given a dataset $S$ containing $n$ points/vectors in $D$-dimensional space, i.e., $S = \{p_1, p_2, \ldots, p_n\}$, where $p_i \in \mathbb{R}^D$, we use the squared Euclidean distance\footnote{Squaring does not affect the order of distances.} to compute the distance $dis(p, q)$ between two points $p$ and $q$, where $dis(p, q) = \|p - q\|^2$. The time complexity of computing $dis(p, q)$ is $O(D)$ by scanning each dimension sequentially.

The problem of \textbf{K Nearest Neighbor (KNN)} search is to find the data points in $S$ that have the top-K smallest distances to a query point $q \in \mathbb{R}^D$.
Due to the complexity of KNN search, a relaxed version of the problem, known as \textbf{Approximate K Nearest Neighbors (AKNN)} search, has been proposed. 
Given a query point $q$, AKNN search allows the returned points to be close to, but not necessarily the exact, K closest points to $q$. 
This approach sacrifices some accuracy in favor of improved computational efficiency.

Note that there are other widely adopted distance metrics, such as cosine similarity and inner product, which can be transformed into Euclidean distance through simple transformations~\cite{ADSampling:journals/sigmod/GaoL23}. 
Therefore, our discussion will focus solely on AKNN search under the Euclidean distance metric. Table~\ref{tab:notation} summarizes the commonly used notations.

\begin{table}[t]% h asks to places the floating element [h]ere.
  \caption{A Summary of Notations}\vgap
  \label{tab:notation}
  \small
  \begin{tabular*}{\linewidth}{@{\extracolsep{\fill}} p{12mm} | p{67mm}}
    \toprule
    Notation   &  Description\\
    \midrule
    $S$        &  A set of points/vectors \\
    $D$        &  The dimensionality of $S$ \\
    $\RR$        &  The projection (rotation) matrix \\   
    $\mathbb{R}^d$ & $d$-dimensional Euclidean space \\
    $dis,dis^{\prime}$ & Exact and approximate distance \\
    %$N^{prob}$  & The search parameter of $\IVF$\\
    %$N^{ef}$    & The search parameter of $\HNSW$ \\
    $\|u,v\|$  &  The Euclidean distance between $u$ and $v$ \\
    $\tau$   & The distance threshold in the queue $Q$\\
    $\epsilon$ & The estimation error\\
    $L$ & The linear classifier\\
    \bottomrule
  \end{tabular*}%\vgap\vspace{-1.5em}
\end{table}\vgap

\stitle{AKNN Algorithms.}
Currently, AKNN algorithms can be mainly divided into four categories: inverted file-based~\cite{IMI:DBLP:journals/pami/BabenkoL15,DBLP:journals/pami/PQJegouDS11}, graph-based~\cite{DBLP:journals/pami/HNSWMalkovY20,DBLP:journals/pami/SSGFuWC22,DBLP:journals/pvldb/NSGFuXWC19,HVS:DBLP:journals/pvldb/LuKXI21,tMRNG:journals/pacmmod/PengCCYX23,FANNG:harwood2016fanng,DPG:DBLP:journals/tkde/LiZSWLZL20}, tree-based~\cite{Random-Project-Tree-2008}, and hash-based~\cite{SRS-yifang-2014,PMLSH-bolong-2020,RPLSH-1995,C2LSH-2012-SIGMOD,QALSH-2015-VLDB,query-aware-LSH,VHP-VLDB-2022,Hmsearch-SSDM-2013,DB_LSH2-TKDE-2023}.

% \begin{figure}[!t]
%     \centering
%     \includegraphics[width=0.90\columnwidth]{figures/HNSW-IVF-demo.pdf}\vgap
%     \caption{The Illustration of $\IVF$ and $\HNSW$}\vgap\vgap\label{fig:IVF-HNSW-demo}
% \end{figure}
%\footnote{The number k of clusters in the k-means algorithm and the number K of neighbors in the AKNN search need not match.} 

\sstitle{Inverted File-Based.}
Inverted file-based algorithms, such as $\IVF$~\cite{IMI:DBLP:journals/pami/BabenkoL15}, are often used to speed up AKNN search. 
The core idea of $\IVF$ is to cluster the points in a data set $S$ into multiple clusters, which helps to speed up the search process. During the \textbf{indexing} phase, $\IVF$ uses the k-means algorithm to cluster the data points in $S$. 
It then constructs a bucket for each cluster and assigns the data points within that cluster to the corresponding bucket. 
In the \textbf{query} phase, given a query point $q$, $\IVF$ first selects the nearest top-$N^{Probe}$ clusters based on the distance from $q$ to the cluster centroids. 
It then retrieves all data points in the corresponding buckets of these nearest clusters as candidates and identifies the K nearest neighbors among these candidates. 
% Here, $N^{Probe}$ is a parameter that balances time and accuracy: increasing $N^{Probe}$ considers more clusters, thereby improving accuracy at the expense of increased search time.

\sstitle{Graph-Based.}
Graph-based algorithms for AKNN search construct a navigable graph where nodes represent data points and edges connect nodes that are nearest neighbors. 
Hierarchical Navigable Small World ($\HNSW$)~\cite{DBLP:journals/pami/HNSWMalkovY20} is a prime example of such algorithms, known for its superior search speed and accuracy.
%(see Fig.~\ref{fig:IVF-HNSW-demo}).
During the \textbf{indexing} phase of $\HNSW$, data points are inserted into a multi-layered graph structure, with each layer representing data at increasingly fine-grained levels. 
Each point is connected to a fixed number of closest neighbors, ensuring each layer maintains a navigable small-world network property.
In the \textbf{query} phase, the search begins from the top layer, leveraging the hierarchical small-world structure to efficiently navigate towards the region closest to the query point. 
Upon reaching the base layer, the algorithm navigates precisely through the neighborhood graph to identify the approximate nearest neighbors to the query point.

In this paper, we only consider graph-based and IVF-based indices due to their outstanding performance and widespread application.

% \subsection{Existing Distance Computation Methods}\label{sub:distance}
% The introduced AKNN algorithms generally follow a common two-phase framework: candidate generation and refinement.

% \stitle{Candidate Generation Phase.}
% In this phase, AKNN algorithms gather a superset of points in set $S$ as candidates. The strategies for generating these candidate points significantly differ among algorithms, thereby forming distinct categories. 
% For instance, $\IVF$ utilizes clustering techniques, whereas $\HNSW$ employs traversal methods to identify candidates.

% \stitle{Refinement Phase.}
% In this phase, AKNN algorithms identify the top-K closest points to a query point $q$ from the generated candidates. 
% Notably, this phase is consistent across different AKNN algorithms.
% To determine the K nearest points from the candidates, these algorithms typically utilize a queue $Q$, commonly structured as a max-heap. 
% They sequentially examine each candidate point. For each candidate point $p$, the algorithm checks whether the distance from $p$ to the query point $q$ is less than or equal to the maximum threshold $\tau$ recorded in $Q$. 
% If this condition is met, the candidate point $p$ is inserted into $Q$, and $Q$ is subsequently updated. 
% If not, $p$ is disregarded.

% \begin{figure}[!t]
% \begin{small}
%     \centering
%     \includegraphics[width=0.80\columnwidth]{figures/appdist-demo.pdf}\label{fig:app-dist}\vgap
%     \caption{\revise{The Example of Projection and Quantization}}\vgap
% \end{small}
% \end{figure}

% \stitle{Approximate Distance Computation.}
\subsection{Existing Distance Computation Methods}\label{sub:distance}
The time to compute distances dominates the runtime of AKNN search, accounting for 80\% of the time complexity in $\IVF$ and 90\% in $\HNSW$. 
To improve efficiency, an intuitive idea is to use approximate distances instead of exact distances for the refinement phase of the AKNN search. 
Two types of methods are proposed for computing approximate distances: projection and quantization. 

%Recall that computing distances by successively scanning dimensions leads to a linear time cost in the dimensionality $D$ of $S$.
%Thus, these methods use different ideas to reduce the dimensionality to improve efficiency, and each type has its own application scenarios.

\sstitle{Projection.}
Projection methods, such as random projection, map high-dimensional data to a lower-dimensional space, mitigating the curse of dimensionality and facilitating efficient data processing and storage. 
Typically, dimensionality reduction is achieved by multiplying the points in the original space by an orthogonal projection matrix. 
The advantage of projection methods is their relative ease of implementation, which allows for the processing of large, high-dimensional datasets in a comparatively short amount of time.

\sstitle{Quantization.}
Quantization methods, such as Product Quantization (PQ) and Residual Quantization (RQ), map the original vector into discrete short quantized codes with a pre-trained codebook. 
Unlike projection methods, which compute distance in low-dimensional space, quantization computes the distance between the query vector and base quantized codes as the approximate distance. Due to its quantized nature and the use of a codebook, quantization can speed up distance computation via distance look-up.
%, making it effective for distance computing.

\begin{comment}
The advantages of product quantization over projection include:

$\bullet$ Storage efficiency. product quantization represents high-dimensional data using a combination of indices from its quantized subspaces, which often requires less memory than storage with projection-based methods; 

$\bullet$ Computational Speed. Due to its quantized nature and the use of a codebook, product quantization can speed up the distance lookup search process, making it effective for distance computing.
\end{comment}

\stitle{Remark.}
Both projection and quantization methods can speed up the computation of distances. 
However, using approximate distances as a direct substitute for exact distances in the refinement phase of ANN algorithms (without distance correction) can result in reduced search accuracy. For example, none of the quantization methods achieve more than 60\% recall without re-ranking~\cite{learning-to-hash-survy-2017,ADSampling:journals/sigmod/GaoL23}.
To illustrate, consider the scenario where K $= 1$ and we want to find the nearest neighbor of a query point $q$. 
If the approximate distance of any candidate point $p$ to query $q$ is less than the approximate distance of the true nearest neighbor of $q$, we can not return an exact result.

\section{Problem Analysis}\label{sec:problem}
To mitigate the loss of accuracy often observed when directly integrating approximate distances in AKNN algorithms, $\ADS$ has been introduced as an optimization technique for distance computations. 
The key idea behind $\ADS$ is not only to leverage approximate distances but also to incorporate an error bound for correction purposes.
This treatment also allows $\ADS$ to determine if the corrected approximate distance is adequate for the refinement phase of the AKNN search. 
When the current approximate distance falls short, $\ADS$ triggers more precise distance computations to compensate for the deficiencies. 
By incorporating these incremental computations for correction, $\ADS$ enhances the overall accuracy of AKNN search.

\stitle{Distance Estimation.}
$\ADS$ first employs random projection to reduce the dimensionality from $D$ to $d$ for points, thereby estimating an approximate distance $dis^{\prime}$ for the exact distance $dis$. 
The relationship between any pair of approximate and exact distances is given by the following lemma:

\begin{lem}[\cite{ADSampling:journals/sigmod/GaoL23}]\label{lem:JL}
    For a given point $x \in \mathbb{R}^D$, a random projection $P\in \mathbb{R}^{d \times D}$ preserves its Euclidean norm with a multiplicative error $\epsilon$ bound with the probability of
    \begin{equation}
        \mathbb{P}\left\{\left|\sqrt{\frac{D}{d}}\|P\mathbf{x}\|-\|\mathbf{x}\|\right|\leq\epsilon\|\mathbf{x}\|\right\}\geq1-2e^{-c_0d\epsilon^2}
    \end{equation}
\end{lem}

From Lemma~\ref{lem:JL}, it follows that the error between the approximate distance $dis^{\prime}$ and the exact distance $dis$ is bounded by $\epsilon \cdot dis$ with a small failure probability ($2e^{-c_0d\epsilon^2}$).

\stitle{Distance Correction.}
$\ADS$ improves the accuracy of AKNN search by leveraging error bounds between approximate and exact distances for correction.
Then, $\ADS$ employs a hypothesis test to determine whether the (corrected) approximate distance between candidate points and the query point is sufficient to exclude candidates during the refinement phase.
That is, if $dis^{\prime} > (1+\epsilon) \cdot \tau$, or $dis^{\prime} - \epsilon \cdot \tau > \tau$, where $\tau$ is the maximum distance (threshold) in the queue $Q$, $\ADS$ infers that $dis > \tau$ at a given significance level $p = 2e^{-c_0\epsilon_0^2}$. 
Here, $\epsilon_0$ is a parameter that requires empirical tuning. 
In this case, excluding candidate points from $Q$ based on the corrected approximate distance $dis^{\prime} - \epsilon \cdot \tau$ is reliable.

Conversely, if the approximate distance does not satisfy the exclusion condition, it is insufficient to determine whether a candidate point should be removed from the queue $Q$. 
In such cases, $\ADS$ requires using additional dimensions to refine the distance estimation. 
This process entails calculating a more accurate approximate distance to decisively conclude whether $dis > \tau$ or $dis \le \tau$.
The correction procedure continues incrementally until all dimensions have been sampled and the final accurate distance is obtained.

\stitle{Limitations.}
While $\ADS$ demonstrates superior performance compared to methods relying solely on approximate distances, it also presents two notable limitations:

\sstitle{(1) Limited Effectiveness.}
$\ADS$ employs random projection to compute approximate distances. However, among various projection methods, random projection does not ensure the minimization of error between approximate and exact distances. 
This discrepancy suggests that the approximate distances derived from random projection may significantly deviate from the exact distances. 
Notably, $\ADS$ requires incremental calculations of approximate distances until it can conclusively determine whether to exclude a candidate point. 
Therefore, finding a way to enhance the accuracy of approximate distance estimation could enable $\ADS$ to stop calculating distances for a candidate point earlier, thus accelerating the computation process.

\sstitle{(2) Lack of Generality.}
The error bound provided by Lemma~\ref{lem:JL} is applicable only to scenarios where the projection matrix is random. 
This limitation underscores the absence of a more general distance correction scheme that can adapt to other approximate distances, such as quantization distances. 
Developing a generalized distance correction scheme could potentially improve the efficiency and applicability of $\ADS$, especially when other approximate distances are shown to be more effective than those derived from random projection~\cite{Learning-Hash-survey-2017-PAMI}.

%% file: Methodology.tex
\section{An Improved Projection-Based Distance Computation}\label{sec:estimation}
This section primarily addresses the first limitation of $\ADS$: the limited effectiveness of distance estimation based on random projection. 
We demonstrate that using optimal orthogonal projection, rather than random projection, leads to more effective distance estimation. 
Additionally, we propose a corresponding distance correction method tailored to this newly developed distance estimation technique.
%In this section, we focus on projection-based distance estimation, while the general distance computation method will be introduced in the next section.

\subsection{Distance Decomposition}
We investigate which projection matrices produce approximate distances that are close to the exact distances. 
To achieve this, we decompose the exact distance into its corresponding approximate distance and the estimation error introduced by the projection or rotation. 
By minimizing this estimation error, we can determine the optimal projection matrix that provides the most accurate projection-based distance estimation. 
The discussion of non-projection-based distance estimation is left to the next section.

%\stitle{Distance Decomposition.}
Let $\xx$ and $\qq$ represent the $D$-dimensional data vectors and query vector, respectively. 
We consider a simple model where data vectors $\xx$ are randomly sampled from the dataset following an unknown fixed distribution $U$. 
To capture global transformations, we introduce a rotation parameterized by the matrix $\RR$.
The transformed (rotated) vectors are denoted as $\xx_D = \RR \xx$ and $\qq_D = \RR \qq$. 
We partition the rotated data vector $\xx_D$ into two components: $\xx_d$, consisting of the first $d$ \textbf{projected dimensions}, and $\xx_r$, which represents the \textbf{residual dimensions}. 
Similarly, we decompose the query vector as $\qq_D = (\qq_d, \qq_r)$.
The exact distance between the transformed vectors can then be expressed as a decomposition:

\begin{equation}
\begin{aligned}
    \|\xx - \qq\|^2 = \|\xx_D - \qq_D\|^2 &= \|\xx\|^2 + \|\qq\|^2 - 2 \cdot \Braket{\qq, \xx} \\
    &= \|\xx_d\|^2 + \|\qq_d\|^2 + \|\xx_r\|^2 + \|\qq_r\|^2 \\
        &-2 \cdot (\Braket{\qq_d, \xx_d} + \Braket{\qq_r,\xx_r}).
\end{aligned}\label{eq:dist-decomp}
\end{equation}

Let $C_1 = \|\xx_d\|^2 + \|\qq_d\|^2 + \|\xx_r\|^2 + \|\qq_r\|^2$ and $C_2 = -2 \cdot \Braket{\qq_d, \xx_d}$.  
For the $C_1$ term, $\|\xx_d\|^2 + \|\xx_r\|^2$ can be precomputed and stored, while $\|\qq_d\|^2 + \|\qq_r\|^2$ only needs to be computed once for each query. 
The $C_2$ term, $-2 \cdot \Braket{\qq_d, \xx_d}$, can be calculated with an $O(d)$ cost. 
Thus, with $O(d)$ computation, the approximate distance can be computed as $dis^{\prime} = C_1 - C_2$, with the \textbf{estimation error} term compared to the exact distance being $\epsilon = -2 \cdot \braket{\qq_r, \xx_r}$.

\subsection{An Improved Distance Estimation}
Equation~\ref{eq:dist-decomp} presents the estimation error, expressed as $\epsilon = -2 \cdot \braket{\qq_r, \xx_r}$, which captures the difference between the exact and approximate distances. 
Assuming the data vector follows a Gaussian distribution\footnote{The data vectors have been centralized to have a mean of zero.}, i.e., $\xx \sim \mathcal{N}(0, \Sigma)$ for a given query $\qq$, the distribution of the estimation error can be viewed as a linear combination of multiple Gaussian distributions.
Under this assumption, we show why random projection yields suboptimal results.

\stitle{Which Projection is Better?}
We aim to determine which projection matrix provides the most accurate distance estimation. 
Under the constraint of orthogonal projection (which preserves distance after rotation), we plan to minimize the variance in estimation error. 
%By this, we can identify the most effective projection that ensure the estimated distance is as close as possible to the exact distance.
%To this end, we first consider the computation of the variance related to the estimation error term. 
Let $\sigma_i^2$ denote the variance of the $i$-th dimension in the distribution $U$ (where the data vectors are assumed to follow the distribution $U$). 
Upon receiving a query, the variance of the inner product for the residual dimension is given by $\sigma_i^2 q_i^2$. 
Therefore, the variance of the estimation error term can be computed as follows:

\begin{equation}\label{eq:Err-Var}
 Var(-2 \cdot \Braket{\qq_r,\xx_r}) = 4 \cdot \sum_{i=d+1}^{i\leq D}(\qq_i\sigma_i)^2    
\end{equation}

We are ready to show that the principal component projection matrix, as opposed to the random projection matrix used in $\ADS$, results in the smallest variance in the estimation error term.

\begin{thm}\label{theorem:PCA}
    Given a set of vectors $S$, the Principal Component Analysis (PCA) projection matrix maximizes the variance along the projected dimensions while simultaneously minimizing the variance in the residual dimensions. This optimization is achieved over all possible orthogonal projection matrices.
\end{thm}

% \begin{proof}
%     Sketch: Considering each set of orthogonal bases $\ww_i$ in the projection matrix satisfies the unit norm constraint $\ww_1^T\ww_1=1$. The optimization objective of PCA is to maximize $\ww_1^T \Sigma \ww_1$, where $\Sigma$ is the covariance matrix. The stationary points of this optimization problem will be the eigenvectors of $\Sigma$, with the corresponding eigenvalues representing the variance. Same as PCA, we sort the eigenvectors according to their eigenvalues; the principal components are the eigenvectors $\ww_1, \ww_2, \ldots, \ww_d$ corresponding to the $d$ largest eigenvalues $\lambda_1, \lambda_2, \ldots, \lambda_d$, and the residual dimension corresponds to the $r$ smallest eigenvalues $\lambda_{d+1}, \lambda_{d+2}, \ldots, \lambda_{r}$, which also minimizes the variance of the residual dimensions.
% \end{proof}

\begin{figure}[t]
	\centering
	\begin{small}
		\subfloat[Comparing PCA and Random]
            {
            \includegraphics[width=0.48\columnwidth]{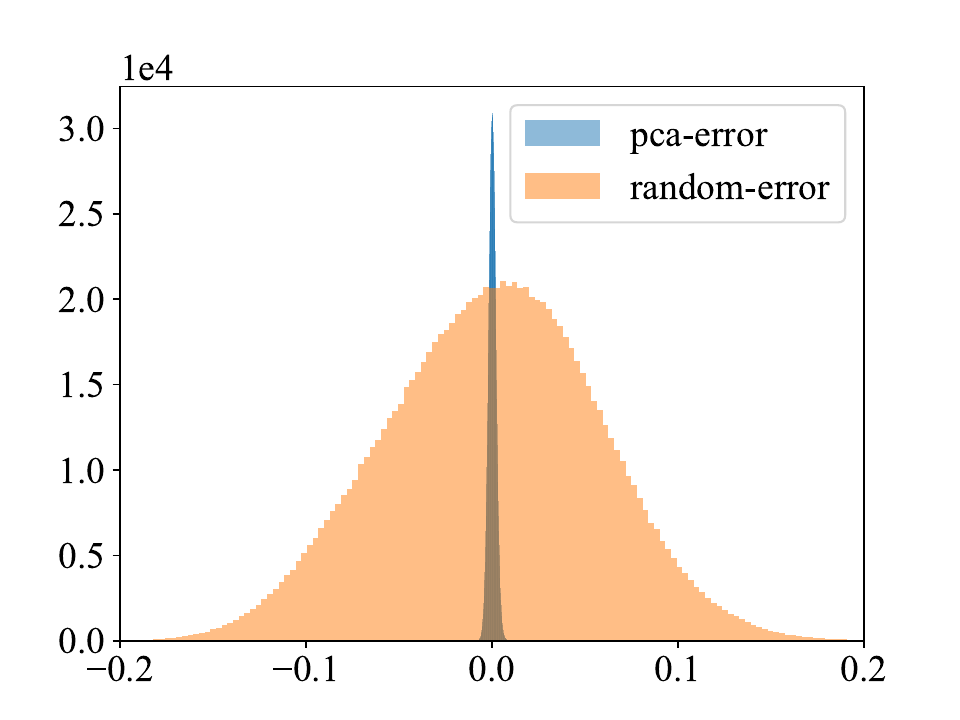}
            %\label{fig:pa}
            }
            \subfloat[Varying dim of PCA Projection]{
            \includegraphics[width=0.48\columnwidth]{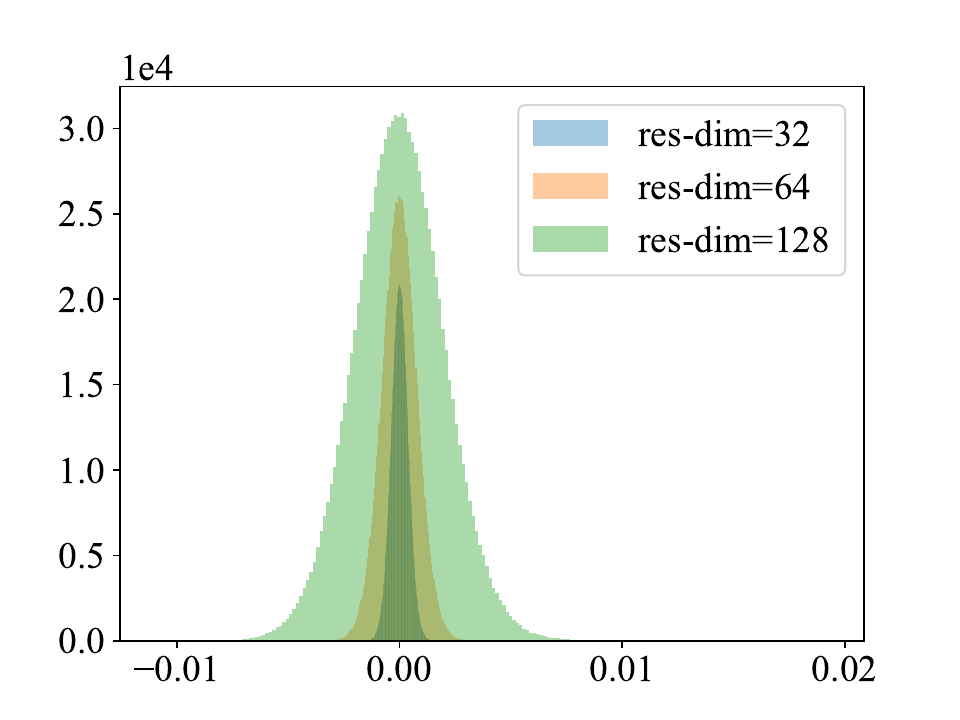}
            %\label{fig:pb}
            }
		\caption{The Distribution of Estimation Error}\label{fig:pca-random-error}
		\vspace{-1mm}
	\end{small}
	\vspace{-1mm}
\end{figure}

The PCA projection matrix is well suited for distance estimation as it maximizes the projection variance, allowing the approximate distances to capture more information and thus become closer to the exact distances. 
Theorem~\ref{theorem:PCA} further shows that PCA projection minimizes the variance in the residual dimensions, effectively reducing the estimation error as much as possible.
To gain deeper insight into Theorem~\ref{theorem:PCA}, we analyze the distribution of the estimation error terms using real data sets and compare different projection matrices.
Specifically, for the DEEP1M dataset (256 dimensions) and a given query $q$, we plot the distribution of $\Braket{\qq_r, \xx_r}$ in Fig.~\ref{fig:pca-random-error}.

As shown in Fig.~1.1, the PCA projection matrix with a residual dimension of 128 has a more concentrated distribution compared to the random projection, which can be attributed to the smaller variance. 
Also, as shown in Fig.~1.2, the error associated with the PCA projection gradually converges to zero as the projection dimensions increase and the residual dimensions decrease. 
This shows that the PCA projection matrix is more effective than the random projection matrix in reducing estimation errors.

\stitle{Remark.}
Equation~\ref{eq:Err-Var} assumes that each dimension of $\xx$ is linearly independent. 
In our implementation, we address this requirement by applying the PCA matrix to align data vectors. 
This treatment reduces the off-diagonal elements of the covariance matrix to zero, ensuring independence among dimensions.

\subsection{An Improved Distance Correction}
The effectiveness of $\ADS$ stems from leveraging Lemma~1, which provides an error bound between exact and approximate distances, for distance correction. 
However, Lemma~\ref{lem:JL} is limited to scenarios where the projection matrix is random, thereby excluding the use of optimal (PCA) projections. 
Fortunately, Equation~\ref{eq:dist-decomp} also quantifies the error between exact and approximate distances, which can be modeled as a random variable. 
We then propose using the error quantile (e.g., the 99.5\% quantile) of this random variable to establish an error bound, which can be efficiently derived from the inverse of its Cumulative Distribution Function (CDF).

\stitle{Error Quantile.} 
%$\ADS$ derives an error bound between exact and approximate distances to determine whether the current approximate distance is sufficient for pruning candidate vectors. 
%However, $\ADS$ requires the projection to be random to compute error bounds, which is not directly applicable to PCA projections.
To establish an error bound for PCA projections, we analyze the distribution of the estimation error, denoted as $\epsilon = dis^{\prime} - dis$. This error can be further expressed as $\epsilon = -2 \cdot \braket{\qq_r, \xx_r}$, as shown in Equation~\ref{eq:dist-decomp}. 
Here, we treat $\epsilon$ as the random variable and note that its distribution follows a Gaussian distribution, i.e., $\epsilon \sim \mathcal{N}(0, \sigma^2)$, where the variance $\sigma^2$ is computed using Equation~\ref{eq:Err-Var}.

Since the error $\epsilon$ follows a Gaussian distribution, the value of $\epsilon$ corresponding to specific quantiles can be calculated. 
This allows us to predefine a probability (i.e., quantile) and determine the corresponding value with a given probability as the error bound. 
For example, by setting the error bound to be three standard deviations from its mean, we can achieve a 99.7\% quantile guarantee, as dictated by the empirical rule for Gaussian distributions. 
Conversely, for a given quantile, the corresponding error bound can be expressed as $m \cdot \sigma$, where $m$ is the multiplier derived from the quantile.

After deriving the error bound, similar to $\ADS$, we use this bound to correct the approximate distance to $dis^{\prime} - m \cdot \sigma$ during the refinement phase. 
Also, if $dis^{\prime} - m \cdot \sigma > \tau$, where $\tau$ is the maximum distance (threshold) for a queue $Q$, we can conclude that $dis > \tau$ with a corresponding quantile/probability related to $m \cdot \sigma$. 
This ensures that excluding candidate points from the queue $Q$ is a reliable decision.
Instead, if the condition fails, excluding a point from the $Q$ queue is not sufficient. 
In such cases, additional dimensions must be sampled for the next round of distance correction.

\begin{figure}[!t]
	\centering
	\begin{small}
		\subfloat[DEEP@dim=32]{
            \includegraphics[width=0.44\columnwidth]{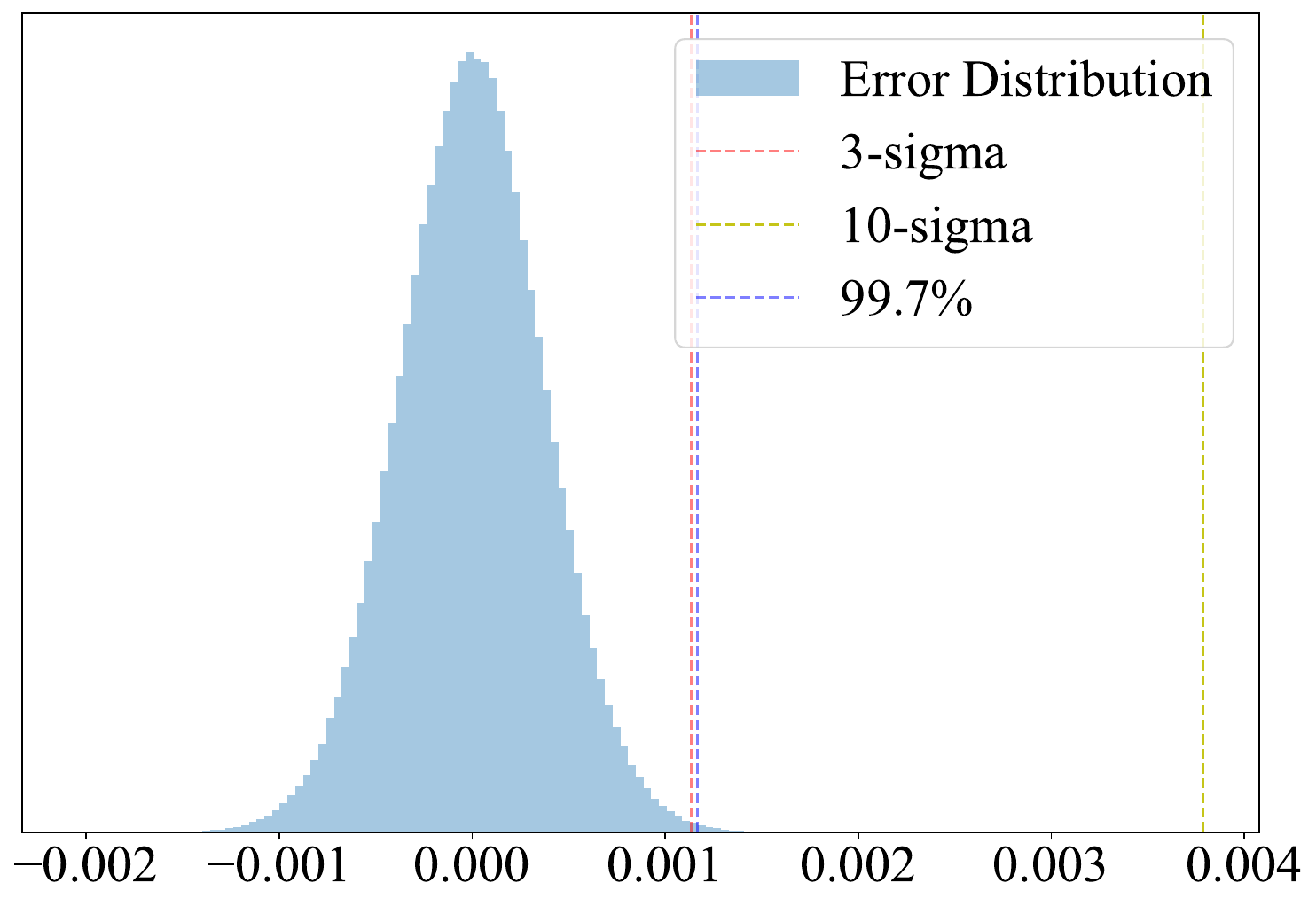}
            }
		\subfloat[DEEP@dim=128]{
            \includegraphics[width=0.44\columnwidth]{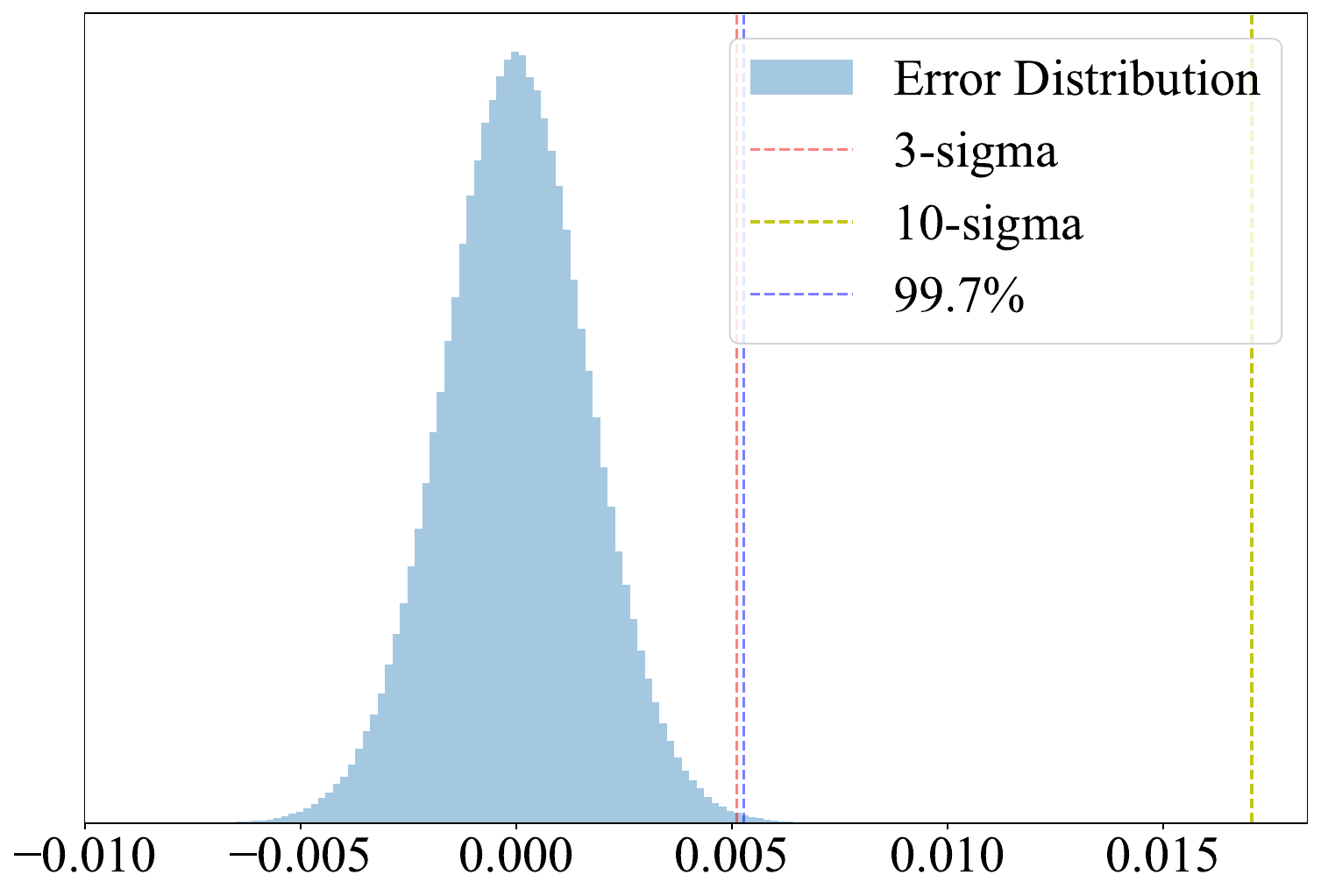}
            }
            \\
            \subfloat[GLOVE@dim=50]{
            \includegraphics[width=0.44\columnwidth]{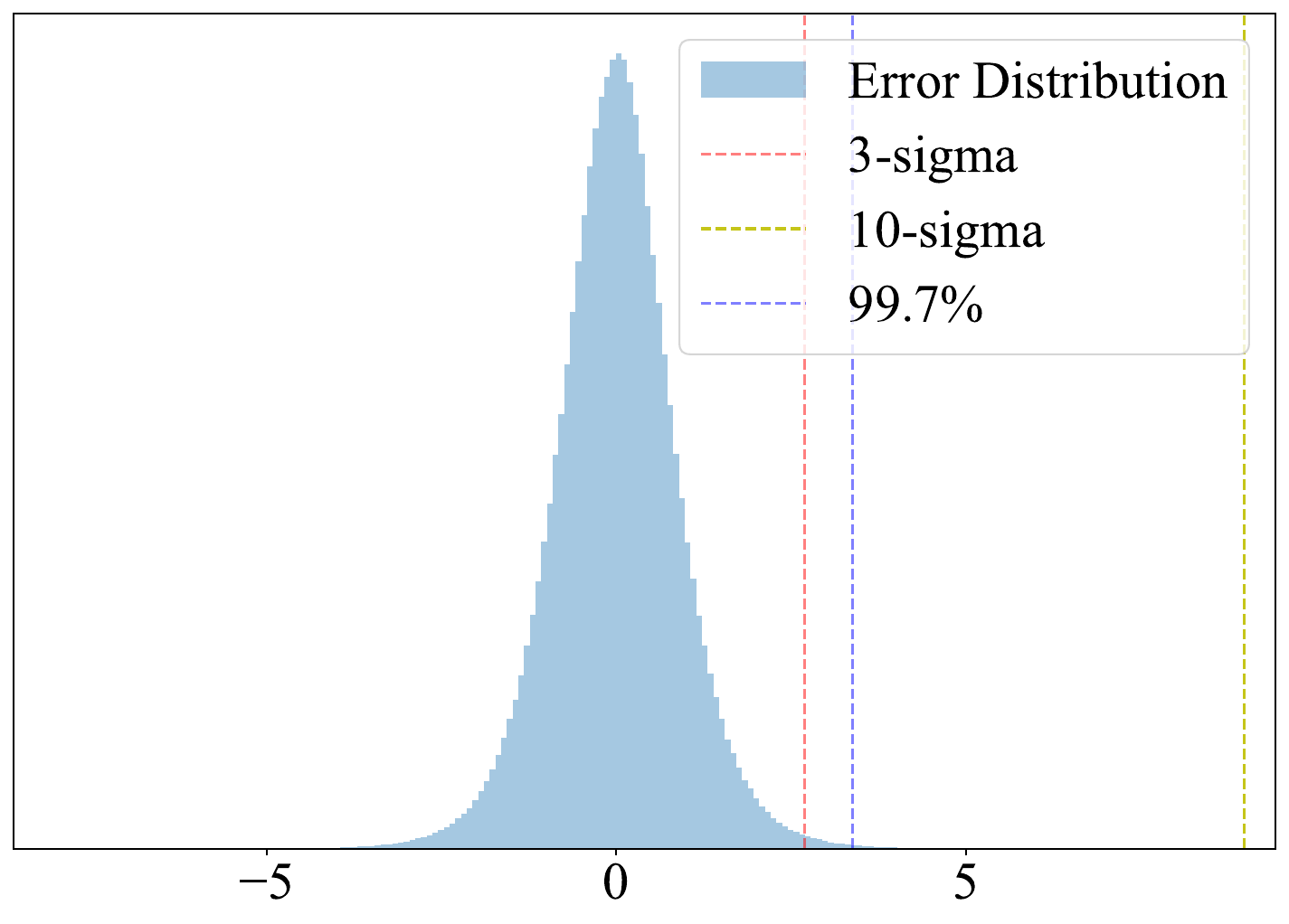}
            }
		\subfloat[GLOVE@dim=100]{
            \includegraphics[width=0.44\columnwidth]{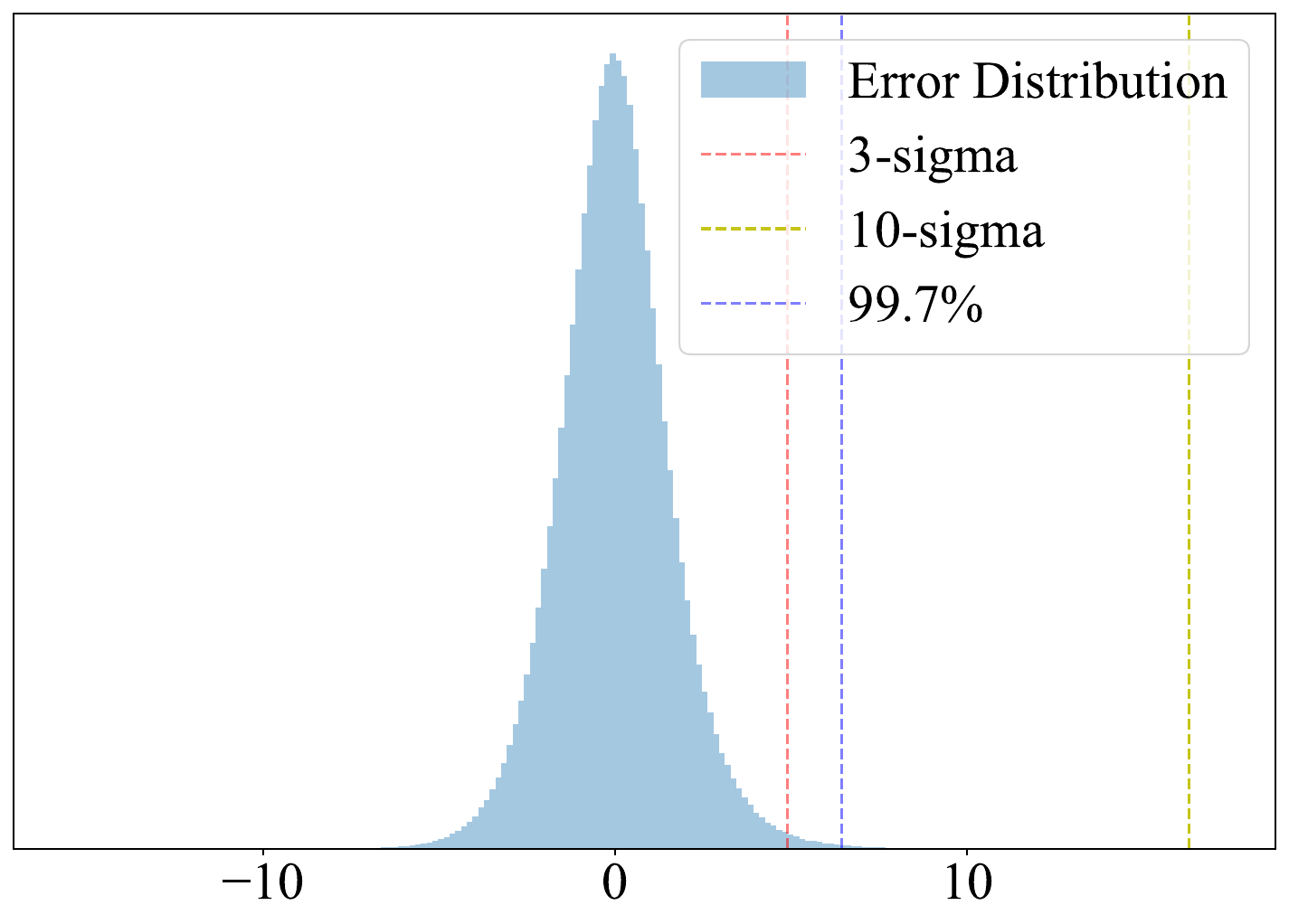}
            }
		\caption{The Empirical Analysis of the New Error Bound}\label{fig:error-distribution-Gaussian-verify-small}
		\vspace{-1mm}
	\end{small}
	\vspace{-1mm}
\end{figure}

%\stitle{Theoretical Analysis.}
However, the PCA projection minimizes the variance of the error, rather than directly minimizing the error corresponding to a given quantile. 
We then prove in Lemma \ref{lem:var=quantile} that minimizing the error variance also minimizes the error quantile. 
This shows that our approach is optimal under the Gaussian distribution, making it highly effective in practice.

\begin{lem}\label{lem:var=quantile}
    Assuming a Gaussian distribution, minimizing the variance of the error also minimizes the corresponding quantile of the error.
\end{lem}
% \begin{proof}
% Sketch: We get $\epsilon\sim\mathcal{N}(0, \sigma^2)$ under the Gaussian distribution assumption. The error quantile can be expressed by the inverse CDF (quantile function) of the Gaussian distribution with probability $p$:
% \begin{equation*}
% F^{-1}(p)=\mu + \sigma\sqrt{2} \mathrm{erf}^{-1} (2p-1)
% \end{equation*}
% The $\mathrm{erf}^{-1}(x)$ is the inverse function of the error function and is positive for $x>0$.
% When the probability $p$ is fixed, $\mu=0$ by centralized, only $\sigma$ affects the value.  Then the variance of the error is minimized, and the right quantile $F^{-1}(p)$ ($p>0.5$) of the error is also minimized.
% \end{proof}

\stitle{Empirical Analysis.}
To gain a deeper understanding of our new error bound, we analyzed real-world datasets and recorded the empirical error distribution for the DEEP and GLOVE datasets. The results, illustrated in Fig.~\ref{fig:error-distribution-Gaussian-verify-small}, show error distributions with projected dimensions set to 32 and 128.
In the DEEP dataset, the empirical error distribution aligns well with the Gaussian distribution: our Gaussian Empirical Rule ($\mu \pm 3\sigma$, depicted by the red line) corresponds closely to the 99.7th percentile of the observed dataset (indicated by the blue line). In contrast, the bound provided by $\ADS$ ($\mu \pm 10\sigma$, represented by the yellow line) diverges significantly from the 99.7th percentile (purple line).
For the GLOVE dataset, despite minor deviations between our bound and the 99.7th percentile, this gap can be effectively managed using the learning-based approach discussed in \S~\ref{sec:correction}.

\subsection{Implementation}
We implement the proposed distance estimation and correction techniques that make up our distance computation method. Also, we propose an optimization to improve this method.

\stitle{A Basic Method.}
Given a query vector $\qq$ and a data vector $\xx$ after the PCA projection, our novel distance computation approach \BSAR is detailed in Algorithm~\ref{algo:BSA-res}.
We begin by calculating $C_1$, defined as $\|\xx\|^2 + \|\qq\|^2$, in Line 1. 
Next, we determine $C_2$ in Line~2, based on the projected dimension $d$.
For the residual dimensions, we pre-compute $\sigma_i^2$ for each data vector along dimension $i$, followed by computing $\sigma$ as the standard deviation of the error (Line~3). 
We then estimate the approximate distance, denoted as $dis' = C_1 - C_2$, and apply a correction using $m \cdot \sigma$.
If the corrected distance exceeds the threshold $\tau$ of the queue, the candidate is pruned, and we return 1 (Line~4-5). If not, we compute the exact distance and return $0$ (Line~6-8).

\begin{algorithm}[!t]
	\caption{\BSAR algorithm}
	\label{algo:BSA-res}
	\begin{small}
	\KwIn{Threshold $\tau$, Multiplier $m$, Project dim $d$, Transformed query $\qq$, Transformed data $\xx$}
	\KwOut{Result: 0 with precise distance $dis$ or 1 with approximate distance $dis^{\prime}$}
	$C_1 \gets \|\xx\|^2 + \|\qq\|^2$; \tcp{Precompute Once}
        $C_2 \gets 2 \cdot \Braket{\xx_d, \qq_d}$; \tcp{Compute On the Fly}
        $\sigma \gets \sqrt{4 \cdot \Braket{\qq_r^2,\sigma_r^2}}$; \tcp{Precompute Once}
	\If{$C_1 - C_2 - m \cdot \sigma > \tau$}{
		\Return{1 with $dis^{\prime}=(C_1 - C_2)$}
	}
	\Else{
            $C_3 \gets 2 \cdot \Braket{\xx_r, \qq_r}$; \tcp{Compute On the Fly}
		\Return{0 with $dis=(C_1-C_2-C_3)$}
	}
\end{small}
\end{algorithm}

\stitle{Optimization.}
One of the advantages of orthogonal projection is its ability to incrementally increase the projected dimensions until all dimensions are used to obtain an exact distance. 
Inspired by $\ADS$, we adopt the incremental correction to form the optimized algorithm in Algorithm~\ref{algo:Multi-BSA-res}.
Specifically, if the current corrected distance, denoted as $C_1 - C_2 - m \cdot \sigma$, is sufficient for pruning, we can return immediately (Line~6–7). Otherwise, we increase the projected dimension by $\Delta_d$ and continue the correction process incrementally (Line~8–9), rather than computing the exact distance directly.

\begin{algorithm}[!t]
	\caption{Incremental-\BSAR}
	\label{algo:Multi-BSA-res}
	\begin{small}
	\KwIn{Threshold $\tau$, Multiplier $m$, Incremental Project dim $\Delta_d$, Transformed query $\qq$, Transformed data $\xx$}
	\KwOut{Result: 0 with precise distance $dis$ or 1 with approximate distance $dis^{\prime}$}
	$C_1 \gets \|\xx\|^2 + \|\qq\|^2$; \tcp{Precompute Once}

        \While{$d<D$}{
            $C_2 \gets C_2 + 2 \cdot \Braket{\xx_{\Delta_d}, \qq_{\Delta_d}}$; \tcp{Incremental}
            $r \gets D-d$\;
            $\sigma \gets \sqrt{4 \cdot \Braket{\qq_r^2,\sigma_r^2}}$\;
    	\If{$C_1 - C_2 - m \cdot \sigma > \tau$}{
    		\Return{1 with $dis^{\prime}=(C_1 - C_2)$}
    	}
    	\Else{
                $d \gets d + \Delta_d$\;
    	}
        }
\end{small}
\end{algorithm}

\begin{figure}[!t]
    \centering
    \includegraphics[width=0.99\columnwidth]{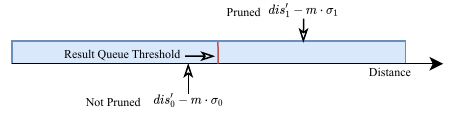}
    \vgap\vgap\vgap\caption{The Example of How Our Methods Work}\vgap\vgap
    \label{fig:incremental}
\end{figure}

\begin{exmp}
Fig.~\ref{fig:incremental} explains how our methods work. 
First, we apply $m \cdot \Delta_0$ to correct the current approximate distance $dis'_0$. If this corrected distance does not exceed the queue threshold $\tau$, it indicates that the candidate cannot be pruned. 
Algorithm~\ref{algo:BSA-res} will then compute the exact distance. In contrast, Algorithm~\ref{algo:Multi-BSA-res} proceeds by incrementing the projected dimension and recomputing the corrected distance as $dis'_1 - m \cdot \delta_1$. 
If this recomputed distance exceeds $\tau$, the candidate can be pruned.   
\end{exmp}

\section{A General Distance Computation}\label{sec:correction}
This section addresses the lack of generality associated with $\ADS$. 
In the previous section, we proposed a new method, \BSAR, by replacing the random projection in $\ADS$ with a PCA projection. 
However, both $\ADS$ and \BSAR are limited to projection-based distance estimation.
Note that other distance estimation methods, such as product quantization, may outperform projection-based methods in certain scenarios. 
To accommodate a broader range of distance estimation techniques, we introduce a novel, data-driven distance correction scheme that is agnostic to the source of the approximate distances. 
This approach also provides a generalized framework for distance computation, extending its applicability beyond projection-based approximation.

\subsection{A Data-Driven Distance Correction}
Distance correction is essential to overcome the loss of accuracy when the approximate distance $dis'$ is used alone.
Specifically, $\ADS$ uses the condition $dis' - \epsilon \cdot \tau > \tau$ to determine whether the candidate point $p$ is unlikely to be added to the queue $Q$, where $dis' - \epsilon \cdot \tau$ is the corrected approximate distance after using the bound $\epsilon \cdot \tau$;
\BSAR similarly uses the condition $dis' - m \cdot \sigma > \tau$ to exclude candidate point $p$, where $dis' - m \cdot \sigma$ is the corrected approximate distance.
For both methods, if the condition is not satisfied, additional dimensions are sampled to compute a refined approximate distance $dis'$ for the next-round correction.
In summary, the conditions $dis' - \epsilon \cdot \tau > \tau$ in $\ADS$ and $dis' - m \cdot \sigma > \tau$ in \BSAR facilitate early stopping, thereby improving the efficiency.
If the conditions are not met, incremental distance correction can still ensure sustained accuracy.
%and prevents serious performance degradation.

\stitle{Data-Driven Distance Correction.}
Both $\ADS$ and \BSAR require the error bounds $\epsilon \cdot \tau$ and $m \cdot \sigma$ for correction, with these bounds working for projection distances. 
This raises a question: how can we determine the error bound for distance correction for an arbitrary approximate distance, which may not originate from a projection?
To address this, our key insight is to treat the error bound as the parameter(s). 
Recall that the purpose of the error bound is to adjust the approximate distance $dis'$ to $(dis' - \text{parameter})$. 
The parameter(s), or error bound, should ensure that $f_{\text{parameter}}(dis') > \tau$ if and only if $dis > \tau$, where $f_{\text{parameter}}(dis')$ represents the corrected approximate distance.
This observation implies that if the exact distance exceeds $\tau$, the corrected approximate distance (i.e., $f_{\text{parameter}}(dis')$) should also exceed $\tau$, effectively pruning irrelevant candidates. 
Thus, identifying the error bound reduces to determining the parameter(s) that satisfy this pruning condition.
To address this, we propose a data-driven approach to calibrate the parameter(s) for distance correction.

\begin{figure}[!t]
    \centering
    \includegraphics[width=0.98\columnwidth]{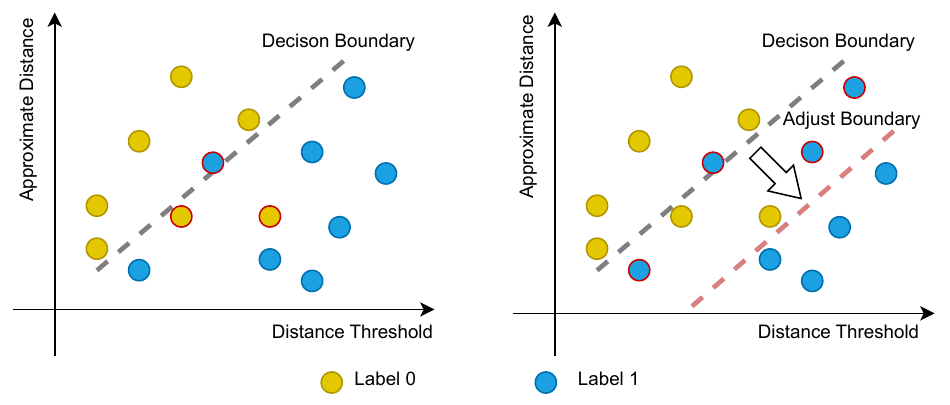}\vgap
    \caption{The Example of the Learned Decision Boundary}\vgap
    \label{fig:decison}
\end{figure}

\stitle{Parameter Learning.}
The next consideration is how to determine the parameter(s).
Recall that our objective is to ensure that $f_{\text{parameter}}(dis') > \tau$ when $dis > \tau$. Equivalently, we need $f_{\text{parameter}}(dis') - \tau > 0$ when $dis > \tau$, and $f_{\text{parameter}}(dis') - \tau \le 0$ otherwise.
If we label the first case (where $dis > \tau$) as label 1, and the second case (where $dis \le \tau$) as label 0, this formulation leads naturally to a classification problem.
Here, the features are $dis^{\prime}$ and the threshold $\tau$, and the task is to obtain the weights of the features for classification.

We implement the above classification model using a linear model, selecting logistic regression with cross-entropy loss trained via stochastic gradient descent (SGD) for its stable performance and high training efficiency.
Empirical results indicate similar outcomes across various linear models, so we omit further discussion on model selection.
Formally, we can express this linear model as:

\begin{equation*}
\begin{aligned}
    L & = sign(w_{1}dis^{\prime} + w_{2}\tau + b > 0) \\
            & = sign(m_{1}dis^{\prime} + \beta > \tau)
\end{aligned}
\end{equation*}
where Label 0: $dis \leq \tau$ and Label 1: $dis > \tau$.

To summarize, we recast the error-bound estimation process as a parameter learning task (where parameters are weights $m_1$ and $\beta$), and we directly use the learned classification model for distance correction, thereby determining whether a candidate point should be excluded or not. 
The training data is sampled from vector points and queries, and the whole process works without any assumptions about the estimated distance $dis'$.

\stitle{Adaptive Adjustment.}
After obtaining the initial parameters, we derive the decision boundary $m_{1} \cdot dis^{\prime} + \beta$.
One advantage of our data-driven correction approach is the flexibility it provides in adjusting the parameter $\beta$ to $\beta^{\prime}$, thereby modifying the decision boundary. 
This adjustment enables a balance between accuracy (specifically recall) and efficiency, as achieving high recall for Label 1 (i.e., when $dis > \tau$) is essential for optimizing efficiency: In extreme scenarios where Label 1 instances are absent, no pruning occurs, potentially impacting performance.

To implement this idea, we set a target recall on the training set and identify the $\beta^{\prime}$ value that aligns with the desired recall level for a given AKNN search accuracy. A binary search on $\beta^{\prime}$ is then conducted to ensure the model achieves the target recall $r$ for Label 0, enabling automatic and adaptive parameter configuration to meet specific recall targets.

\begin{exmp}
    The left figure in Fig.~4 shows the decision boundary of our data-driven approach, with parameters $m_1$ and $\beta$ learned through training data. 
    Data points with red borders indicate misclassified instances. 
    The right figure in Fig.~4 shows how adjusting the boundary by updating $\beta$ to $\beta'$ can affect classification results. 
    This adjustment is designed to trade off a small amount of accuracy for label 1 to ensure that the accuracy for label 0 meets essential performance requirements (i.e., recall targets).
\end{exmp}

\stitle{Remark.}
The data-driven approach can be aligned with \BSAR presented in \S~\ref{sec:estimation} by setting $m_1 = 1$ and $\beta = m \cdot \sigma$. 
Also, this approach offers greater flexibility as it can accommodate any approximate distance, enhancing its generality.

\subsection{Implementation}
This section provides details on the implementation of our proposed data-driven distance correction. We also show how to implement incremental distance correction.

\stitle{Approximate Distances.}  
Our new distance correction method is designed to be flexible across various types of approximate distances.
For projection distances, we use a straightforward PCA projection as an approximate distance measure without applying the decomposition given by Equation~\ref{eq:dist-decomp}. 
This approach is referred to as \BSAP. 
The features of the model include the approximate distance and the threshold.

For quantization distances, we calculate the distance from quantized centroids of query $q$ to the data point $u$, known as the asymmetric distance, as the approximate distance $dis^{\prime}$. 
We also use OPQ~\cite{DBLP:journals/pami/OPQGeHK014} as our quantization distance method, and thus form the final distance computation method denoted as \BSAO. 
Besides the approximate distance and threshold, we incorporate the distance from $u$ to its quantized centroid as an additional feature. 
This additional feature further enhances the effectiveness of the linear model.
Also, we forgo hardware-specific optimizations, concentrating instead on the core distance computation algorithm itself~\cite{PQ-fast-scan-VLDB-2015,DBLP:journals/pami/PQJegouDS11}.

\stitle{Incremental Correction.}
We now discuss the incremental approach for data-driven distance correction. 
Similar to Algorithm~\ref{algo:Multi-BSA-res}, we start by training an initial classifier, after the initial approximate distance is given. 
Each time the classifier fails to confirm that $dis > \tau$ (i.e., it assigns label 0), we incrementally sample additional dimensions to compute a refined approximate distance, $dis$, and train a new classifier.
This process is repeated until the projected dimension matches the data dimension, at which point we obtain an exact distance.

% It is important to note that we do not use multiple linear classifiers with \BSAO. This is because the PQ method cannot perform incremental calculations based on previous approximate distances as the projection method does.
\subsection{Discussion of Out-of-Distribution Query}

For our proposed methods, we need to obtain the error distribution to establish an error bound for \BSAR and to form the training data of the linear models for \BSAO and \BSAP. 
However, in real-world applications, query diversity may cause the error distribution to deviate from the original one. 
Such out-of-distribution (OOD) queries pose a challenge to all of our methods.
We analyze the impact of OOD queries on each method as follows:
First, the \BSAR algorithm treats the query as a deterministic variable when computing the error bound, making it less sensitive to OOD queries. We validate the robustness of \BSAR through empirical studies (see Exp-A.2 in our technical report~\cite{technicalreport}).
Second, for the linear model-based methods (\BSAO and \BSAP), OOD queries have a significant impact because they rely on the query to generate training data. Our experimental studies also confirm this sensitivity (see Exp-A.2 of~\cite{technicalreport}).
To address this limitation, we propose to retrain the model with approximately $100$ OOD queries. 
Experimental results in Exp-A.3 of~\cite{technicalreport} show that this approach effectively mitigates the performance degradation.

%% file: analysis.tex
\section{Analysis of Proposed Methods}\label{sec:analysis}
This section presents an empirical analysis of our distance computation methods. 
This discussion is divided based on the approximate distance employed in the computation.

\subsection{Analysis Under Projection Distance}
Orthogonal projection is flexible in incrementally adding dimensions during distance approximation.
This process continues either until the exact distance is calculated (using all dimensions) or until an early termination is reached based on an error bound to exclude candidate points. 
Thus, the average number of scanned dimensions determines the time cost of our method when using projection-based approximate distances.
In Exp-6 of \S~\ref{sec:experiment}, we measured the average number of dimensions used and observed that our methods require scanning only a small number of dimensions. This result demonstrates the efficiency of our method.

Moreover, the \BSAR and \BSAP methods incur additional time costs due to the need to rotate the query vectors. 
For a single query, the time complexity of performing matrix multiplication for query projection (rotation) is $O(D^2)$. However, this cost is negligible, as experiments (in Exp-3 of \S~\ref{sec:experiment}) confirm that rotation accounts for only 3\% of the time in high-recall AKNN search scenarios.

\subsection{Analysis Under Quantization Distance}
For our method \BSAO using quantization distance, we evaluate the rate of pruned candidate points (i.e., pruned rate) to assess its efficiency. 
Experimental results in Exp-6 show that \BSAO maintains a high pruned rate, indicating its high efficiency.
The quantization method incurs additional computational cost mainly due to the OPQ (Optimized Product Quantization) rotation, which has a time cost of $O(D^2)$, and the construction of a lookup table, which has a time cost of $O(D \cdot 2^{nbit})$, since in the OPQ process $D$ is divided into $m$ subspaces, each containing $2^{nbit}$ quantized centroids.
However, by leveraging the lookup table, asymmetric distances can be computed with only $m$ table lookups, significantly reducing computational requirements. 

Moreover, these methods entail extra storage, specifically requiring $n \cdot m \cdot nbit$ bits to store the quantized representations across the $m$ subspaces. In typical configurations, $m$ is set to a fraction, such as $1/4$ or lower, of the original dimension $D$. Consequently, this setup results in an additional storage cost of approximately $1/32$ of the dataset size when using float32 vectors.

% \subsection{Approximate Distance Statistic}

% \input{figures/r-error}

%% file: exp.tex
\section{Experiments}\label{sec:experiment}
\subsection{Experimental Settings}
\stitle{Datasets.} 
We utilize eight publicly available datasets of varying scales and sources, as detailed in Table~\ref{tab:dataset_details}. 
These datasets have been widely adopted as benchmarks for assessing AKNN algorithms, encompassing both data and query vectors. 
For datasets that provide designated training data, such as GIST and DEEP, we employ this pre-defined training data directly. 
For datasets without designated training data, we randomly sample from the data vectors as the query vectors to form a training set, then remove these samples from the dataset to ensure a clean evaluation set. 
All data in our experiments are stored in the float32 format.

\begin{table}[t!] 
\centering 
\caption{The Description of Datasets} 
\label{tab:dataset_details} 
\begin{footnotesize}
\begin{tabular}{c|c c c c} 
\toprule
\textbf{Dataset}& \textbf{Dimension} & \textbf{Size} & \textbf{Query Size} & \textbf{Type} \\ 
\midrule
MSONG & 420 & 992.272 & 200 & Audio \\
GIST & 960 & 1,000,000 & 1000 & Image\\ 
DEEP & 256 & 1,000,000 & 1000 & Image\\
WORD2VEC & 300 & 1,000,000 & 1000 & Text\\
GLOVE & 300 & 2,196,017 & 1000 & Text\\ 
TINY5M & 384 & 5,000,000 & 1000 & Image\\
TINY80M & 150 & 79,302,017 & 1000 & Image\\
SIFT100M & 128 & 100,000,000 & 1000 & Image\\
\bottomrule
\end{tabular} 
\end{footnotesize}
\end{table}

%%%%%%%%%%%%
\input{figures/result_plot_new}
%%%%%%%%%%%
\stitle{Evaluation Metrics.} 
To assess search accuracy, we use recall@K, defined as $\frac{|T \cap G|}{K}$, where $G$ represents the ground-truth KNN set for a given query within dataset $S$, and $T$ denotes the result set obtained by AKNN algorithms.
For efficiency evaluation, we employ queries-per-second (QPS), which is the number of queries processed per second, including the end-to-end query time. Additionally, for projection-based distance computation methods, we measure the total number of dimensions scanned; for quantization-based methods, we evaluate efficiency using the pruned rate. All metrics are reported as averages over the entire query set.

\stitle{Algorithms.}
Our distance computation methods include \BSAR (see \S~\ref{sec:estimation}.D), \BSAP (see \S~\ref{sec:correction}.B), and \BSAO (see \S~\ref{sec:correction}.B). 
We integrate these methods into AKNN algorithms to form new variants, and we also compare the original AKNN algorithms. 
All tested algorithms include:

\begin{itemize}[leftmargin=2\labelsep]
    \item $\HNSW$: $\HNSW$ with all exact distances computed;
    \item $\HNSW$++: $\HNSW$ with $\ADS$ for distance computing;
    \item $\HNSW$-\BSAO: $\HNSW$ with \BSAO for distance computing;
    \item $\HNSW$-\BSAP: $\HNSW$ with \BSAP for distance computing;
    \item $\HNSW$-\BSAR: $\HNSW$ with \BSAR for distance computing;
    \item $\IVF$: $\IVF$ with all exact distances computed;
    \item $\IVF$++: $\IVF$ with $\ADS$ for distance computing;
    \item $\IVF$-\BSAO: $\IVF$ with \BSAO for distance computing;
    \item $\IVF$-\BSAP: $\IVF$ with \BSAP for distance computing;
    \item $\IVF$-\BSAR: $\IVF$ with \BSAR for distance computing;
    \item FINGER: $\HNSW$ with FINGER for distance computing.
\end{itemize}

\stitle{Implementation Details.} 
All C++ code was compiled using g++ version 11.4.0 with -O3 optimization, and all SIMD operations were disabled, similar to the setup for $\ADS$. Experiments were conducted on an Intel(R) Xeon(R) Platinum 8352V CPU @2.10GHz with 512GB memory, running on Ubuntu Linux.
We first provide details on the AKNN algorithms used. 
For $\HNSW$, the parameter $M$ specifies the number of connected neighbors, while $efConstruction$ controls the quality of AKNN. Following~\cite{DBLP:journals/pami/HNSWMalkovY20}, we set $M=16$ and $efConstruction=500$.
For $\IVF$, as advised in the Faiss library~\cite{faiss:johnson2019billion}, we set the number of clusters to $4,096$.

To implement our data-driven distance computation methods, we use a simple labeling approach during model training. 
Specifically, we select 10,000 vectors from training vectors as training queries.
Then, the KNNs of each training query are assigned as positive samples (label 0). For negative samples (label 1), we collect 500,000 items through a query process.
To train the linear classifier, we precompute approximate distances, thresholds, and additional features offline. 
The classifier is then trained using Binary Cross-Entropy (BCE) loss. 
We also set a recall target $r$ of 0.995 for the time-accuracy trade-off experiment. 
For further implementation details, please refer to our technical report~\cite{technicalreport}. The code is available at github.com/mingyu-hkustgz/Res-Infer.

% \stitle{Approximate Distance Configuration.}
% For the random projection approach, we set $\epsilon_0=2.1$ and $\Delta_d=32$ which is recommended as the best performance in $\ADS$. For \BSAR and \BSAP, we also set every $\Delta_d=32$ dimension to achieve the same condition as $\ADS$. For the \BSAO approach, we set the subspace number as $d/8$ for the GIST dataset and $d/4$ for the others since all the dataset dimensions can be divided by 4. The $nbits$ for \BSAO is set as 8, the default parameter. The target recall is set as 0.995 for the \BSAP and \BSAO methods. For the multiplier $m$ for \BSAR, we set it as 8 for SIFT, GIST, and DEEP, 12 for TINY and WORD2VEC, and 16 for the GlOVE dataset. For the case of multiple classifiers, we set the target recall for each classifier based on $\Delta_d$ as $r_i = (1 - (1-r)/(D/\Delta_d))$.

% \textbf{Search Parameters.} 
% \begin{table}[htbp] 
% \centering 
% \caption{Parameters Statistics} 
% \label{tab:parameters_details} 
% \begin{tabular}{c|c c c c} 
% \toprule
% \textbf{Dataset} & \textbf{subspaces} & \textbf{$\sigma$-multiplier} & \textbf{target recall} \\ 
% \midrule
% SIFT & 32 & 8 & 0.995 \\  
% GIST & 120 & 8 & 0.995\\ 
% DEEP & 64 & 8 & 0.995 \\
% GLOVE & 75 & 16 & 0.995 \\ 
% TINY5M & 96 & 12 & 0.995 \\
% WORD2VEC & 75 & 12 & 0.995\\
% \bottomrule
% \end{tabular} 
% \end{table}
% Target recall is used for training a linear classifier with a recall constraint on label 0 data($k$-NN).

\input{figures/target_recall}

\subsection{Experimental Results}
\stitle{Exp-1: Performance Test.}
To investigate the tradeoff between time and accuracy for various methods, we vary $N^{ef}$ for $\HNSW$, $\HNSW$++, $\HNSW$-\BSAO, $\HNSW$-\BSAP, and $\HNSW$-\BSAR, as well as $N^{probe}$ for $\IVF$, $\IVF$++, $\IVF$-\BSAO, $\IVF$-\BSAP, and $\IVF$-\BSAR. 
Fig.~\ref{fig:naive-time-acc-trade} shows the time-accuracy curve for all algorithms, where the upper right indicates better performance.
Our findings reveal the following: 
(1) Our distance computation methods demonstrate high efficiency. Specifically, our method, \BSAR, achieves a 2x speedup over $\HNSW$ and a 1.45x speedup over $\ADS$ when applied to the TINY80M dataset in combination with $\HNSW$.
Notably, the computation of PAC or OPQ matrix on the database, along with the training of linear models, remains unaffected by data scale due to the sampling strategy employed. 
Specifically, for large datasets, following the empirical findings of the Faiss library~\cite{faiss:johnson2019billion}, we sample 1 million data points from the database to derive the PCA matrix; we sample 65,536 data points to obtain the OPQ matrix.
%These results underscore the superiority of our distance computation methods, especially when integrated with existing AKNN algorithms.
(2) Our PCA-based distance computation methods, \BSAR\ and \BSAP, when used with AKNN algorithms such as \HNSW, typically outperform quantization-based methods such as \BSAO\ on image datasets, including GIST and SIFT. 
This advantage is due to the skewed variance in these datasets. 
For example, a PCA projection to 32 dimensions preserves 67\% of the variance in the GIST dataset and 82\% in the SIFT dataset.
In contrast, \BSAO\ outperforms PCA-based methods on datasets such as WORD2VEC and GLOVE. In these datasets, the variance is more evenly distributed, with a 32-dimensional PCA retaining only 36\% and 18\% of the variance for WORD2VEC and GLOVE, respectively. This observation suggests that analysis of variance skewness can effectively guide the selection of our proposed methods.

%We also adapt the split result queue strategy in $\HNSW$++. The $\IVF$++ denotes the $\ADS$ method with cache level optimization, and $\IVF$-\BSAP represents the PCA approximate method with the same optimization as $\IVF$++. The $\IVF$-\BSAO method represents the OPQ approximate method without any cache-level optimization.

\stitle{Exp-2: Varying Target Recall.}
The target recall is used to train the linear model by adaptively adjusting the decision boundary (see \S~\ref{sec:correction}). We examine the effect of target recall on the data-driven methods \BSAO and \BSAP when applied to the AKNN algorithm $\HNSW$ and present results on the GIST and DEEP datasets in Fig.~\ref{fig:target_recall}.
We observe that when the target recall is set to $0.995$, the search algorithm achieves the best trade-off between efficiency and recall loss (less than 0.5\%). Thus, we select this value as the default target recall.

\begin{figure}[!t]
	\centering
	\begin{small}
		\subfloat[Pre-Processing Time]{
            \includegraphics[width=0.49\columnwidth]{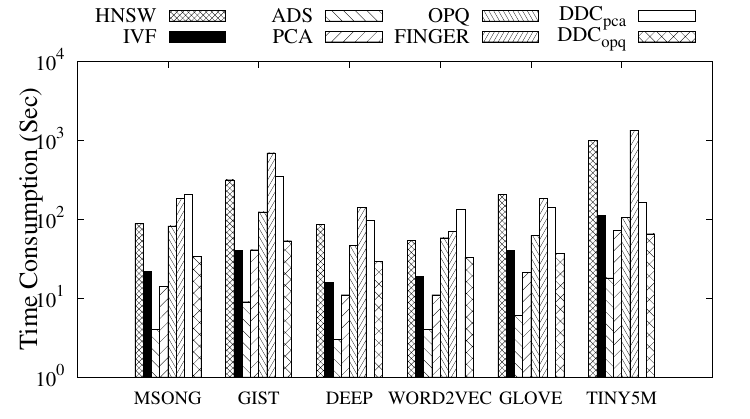}
            }
		\subfloat[Pre-Processing Space]{\hspace{-4mm}
            \includegraphics[width=0.49\columnwidth]{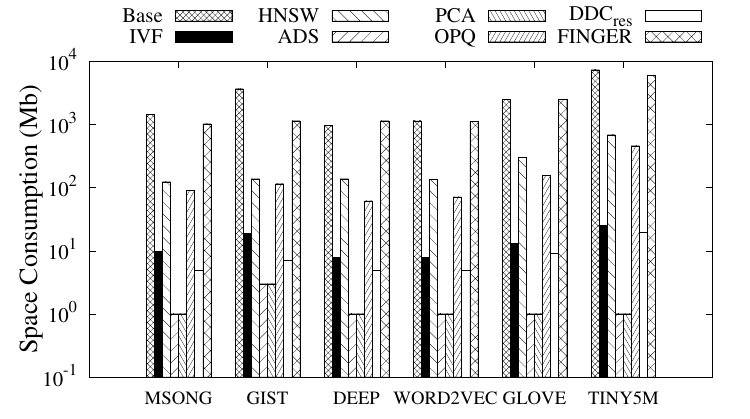}
            }
		\caption{The Test of the Pre-Processing Time and Space}\label{fig:index-time-space}
	\end{small}\vgap\vgap
\end{figure}

\stitle{Exp-3: Test of Pre-Processing Time and Space.}
As introduced in \S~\ref{sec:analysis}, distance computation methods require varying levels of pre-processing time and space. Specifically, $\ADS$ and \BSAR incur additional time and space costs due to projection, while \BSAP and \BSAO, require linear classifier training.
For comparison, we also include $\HNSW$ and $\IVF$ as they need pre-processing time and space for indexing. 
We evaluate these pre-processing costs across different methods, with the results on various datasets presented in Fig.~\ref{fig:index-time-space}.

Fig.~\ref{fig:index-time-space} shows that both $\ADS$ and PCA exhibit low pre-processing times compared to the indexing times of $\HNSW$ and $\IVF$.
In contrast, our methods \BSAP and \BSAO demand more pre-processing time as they involve training a model. Yet, this time remains comparable to the indexing times of $\HNSW$ and $\IVF$.
Moreover, we note that the space required by \BSAP and \ADS, which is solely for the projection matrix ($D^2$ floats), is negligible when compared to the index sizes of $\HNSW$ and $\IVF$. 
Also, \BSAR requires storage for the norms of $N$ vectors, while \BSAO needs to store quantized vectors, which increase linearly with the dataset size. Nonetheless, this space requirement is still comparable to the index space of $\HNSW$ and $\IVF$.

%The ratio between the extra time and the full query time is illustrated in Table~\ref{tab:extra-time-ratio}.

% \begin{table}[!t]
% 	\centering
% 	%\begin{subtable}{1.0\linewidth}
% 		\centering
% 		\begin{tabular}{c|c c c} 
% 			\toprule
% 			\textbf{$N^{ef}$} &\textbf{Proj} & \textbf{PCA} & \textbf{OPQ} \\ 
% 			\midrule
% 			500 &  6.0\% & 8.9\% & 12\% + 5.9\% \\ 
	
% 			1000 & 3.9\% & 5.6\% & 7.6\% + 3.8\% \\ 
	
% 			  1500 & 3.0\% & 3.1\% & 5.8\% + 2.9\% \\

%                 2000 & 2.5\% & 2.9\% & 4.7\% + 2.3\% \\
                
% 			\bottomrule
% 		\end{tabular}
% 	%\end{subtable}
% 	% \\
% 	% \begin{subtable}{1.0\linewidth}
% 	% 	\centering
% 	% 	\begin{tabular}{c|c c c} 
% 	% 		\toprule
% 	% 		\textbf{$N^{probe}$} &\textbf{Proj} & \textbf{PCA} & \textbf{OPQ}  \\ 
% 	% 		\midrule
% 	% 		  100 & 2.5\% & 2.6\% & 3.9\% + 1.9\% \\ 
			
% 	% 		200 & 1.8\% & 1.8\% & 3.0\% + 1.5\% \\ 
			
% 	% 		300 & 1.4\% & 1.4\% & 2.4\% + 1.2\% \\ 
   
%  %                400 & 1.2\% & 1.2\% & 2.1\% + 1.0\% \\ 
% 	% 		\bottomrule
% 	% 	\end{tabular} 
% 	% 	\caption{GIST-IVF}
% 	% \end{subtable}
% 	\caption{Extra Time Ratio on GIST-HNSW}
% 	\label{tab:extra-time-ratio} 
% \end{table}

\stitle{Exp-4: Comparison with FINGER.}
We also compare our methods with FINGER in the same setting. Since FINGER operates exclusively with $\HNSW$, we focus on evaluating $\HNSW$ under various distance computation methods. We present the results on the GIST and DEEP datasets in Fig.~\ref{fig:avx512-time-acc-trade-small}, and the results on other datasets are available in our technical report~\cite{technicalreport}.
The results demonstrate that our \BSAR method is 20\% to 30\% faster than FINGER. 
Additionally, Fig.~\ref{fig:index-time-space} in Exp-3 shows that FINGER requires much more pre-processing time and memory than our method, limiting its feasibility in memory-constrained settings.

\input{figures/result_plot_avx_small}

\begin{figure}[!t]
	\centering
	\begin{small}
        \includegraphics[width=0.58\columnwidth]{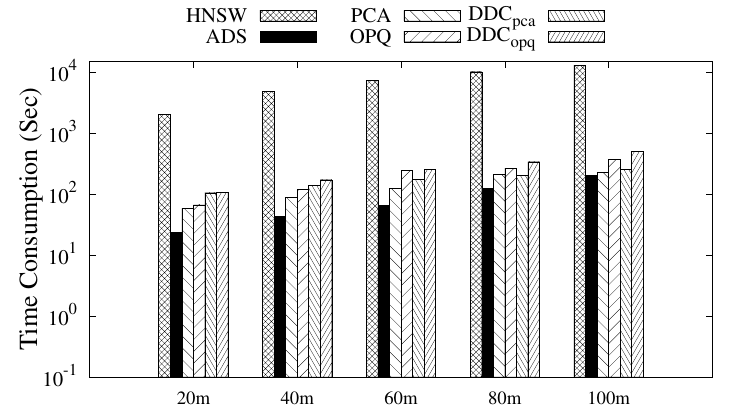}
	\caption{The Test of Scalability on the SIFT100M dataset}\vgap\vgap\label{fig:scale-time}
	\end{small}\vgap
\end{figure}

\stitle{Exp-5: Scalability Test.}
To evaluate the scalability of our methods, we conducted a test on the SIFT100M dataset. 
This dataset was divided into five groups, each containing 20 million, 40 million, 60 million, 80 million, and 100 million entries. 
We used $\HNSW$ as the underlying AKNN algorithm.
In Fig.~\ref{fig:scale-time}, we present the time required by $\HNSW$ to construct the index, as well as the distance computation time for other methods.
We first observe that the pre-processing time for distance methods, such as $\ADS$, PCA, and OPQ, accounts for only 1\%-5\% of the indexing time of $\HNSW$, regardless of dataset size. Our methods, \BSAP and \BSAO, also show small pre-processing time compared to the indexing time of $\HNSW$.
Also, we observe that the training time of linear models for \BSAP and \BSAO increases linearly with dataset size. 
For example, \BSAP and \BSAO require 105 seconds and 107 seconds, respectively, for pre-processing with a dataset of 20 million vectors, and 254 seconds and 504 seconds for a dataset of 100 million vectors.
%This indicates that our methods can improve search efficiency at a cost of less than 10\% of the index construction time.

\stitle{Exp-6: Empirical Analysis of Our Methods.}
In \S~VI, we proposed using the scan dimension and pruned rate metrics to analyze the performance of projection-based methods (\BSAR and \BSAP), and the quantization-based method (\BSAO), respectively. 
We thus compute the average scan dimension ratio (relative to the entire dimension) for various methods on the GIST and DEEP datasets, shown in the left panels of Fig.~\ref{fig:scan-rate-prune-rate}. 
Our methods show superior performance compared to baselines.
For instance, when $N^{\text{ef}} = 2000$, \BSAR scans only 7\% of the total dimensions, \BSAP scans 15\%, and ADsampling scans 26\% on the GIST dataset.
Similarly, we compute the average pruned rate in the right panels of Fig.~\ref{fig:scan-rate-prune-rate}. The results show that our methods achieve a higher pruned rate, further confirming their superiority.

\input{figures/scan-dim-prune-rate}

\stitle{Exp-7: Test of Approximation Accuracy.}
To evaluate the accuracy of our distance computation method, we directly apply our method, \BSAR, to scan the points in the database, without relying on existing AKNN algorithms for finding AKNN. 
Table~\ref{tab:dis-acc@100} presents the accuracy of different methods when projecting to 32 dimensions.
The results indicate that \BSAR achieves higher accuracy than random projection (denoted as Rand) and outperforms PCA across most datasets.
These findings confirm the superiority of our method.

\stitle{Exp-8: Applications in Ant Group.}
Our proposed methods have been applied in the security search applications of Ant Group. 
Specifically, facial images are widely used in digital payment, where AKNN search plays a critical role in security during transactions. 
We conducted experiments on a private dataset from Ant Group, containing 1 million images with 512-dimensional embeddings. Our results show that the proposed \BSAO method reduces retrieval time by 35\% and increases throughput by 55.25\% without sacrificing accuracy, greatly enhancing the efficiency of related applications.
%Also, our \BSAO methd is also used during the re-ranking process in DiskANN~\cite{DiskANN-NIPS-2019} to prune distance computation and disk IO and it improved search efficiency by 50\% under at 90\% recall on the SIFT dataset.

\begin{table}[!t]
    \centering
    \begin{footnotesize}
    \caption{The Test of Approximation Accuracy (Recall@100)}\label{tab:dis-acc@100}
    \begin{tabular}{l|ccc}
    \toprule
        Dataset & PCA & Rand & \BSAR \\
    \midrule
        DEEP & 44.5 & 16.6 & 46.6 \\
        GIST & 34.3 & 7.3  & 51.6 \\ 
        TINY & 32.5  & 7.8 & 43.5 \\ 
        GLOVE & 7.1 & 4.6 & 41.7 \\ 
        WORD2VEC & 18.6 & 8.4  & 29.0 \\ 
    \bottomrule
    \end{tabular}\vgap
    \end{footnotesize}
\end{table}

%% file: figures/target_recall.tex
\begin{figure}[t!]
\centering
\begin{small}
\begin{tikzpicture}
    \begin{customlegend}[legend columns=3,
        legend entries={$\tau=0.9$,$\tau=0.95$,$\tau=0.97$,$\tau=0.99$,$\tau=0.995$,$\tau=0.999$},
        legend style={at={(0.45,1.15)},anchor=north,draw=none,font=\scriptsize,column sep=0.1cm}]
    \addlegendimage{line width=0.15mm,color=red,mark=o,mark size=0.3mm}
    \addlegendimage{line width=0.15mm,color=orange,mark=triangle,mark size=0.3mm}
    \addlegendimage{line width=0.15mm,color=forestgreen,mark=square,mark size=0.3mm}
    \addlegendimage{line width=0.15mm,color=pink,mark=star,mark size=0.3mm}
    \addlegendimage{line width=0.15mm,color=navy,mark=oplus,mark size=0.3mm}
    \addlegendimage{line width=0.15mm,color=violate,mark=otimes,mark size=0.3mm}
    \end{customlegend}
\end{tikzpicture}
\\[-\lineskip]
\vspace{2mm}
% gist@20
\subfloat[GIST ($\HNSW$-\BSAO)]{

\begin{tikzpicture}[scale=1]
\begin{axis}[
    height=\columnwidth/2.75,
width=\columnwidth/2.0,
xlabel=recall@20,
ylabel=Qpsx10,
label style={font=\scriptsize},
tick label style={font=\scriptsize},
% ymajorgrids=true,
% xmajorgrids=true,
% grid style=dashed,
]
\addplot[line width=0.15mm,color=red,mark=o,mark size=0.3mm]%hnsw-pca++ gist
plot coordinates {
    ( 0.91, 41.273 )
    ( 0.95, 27.212 )
    ( 0.961, 20.826 )
    ( 0.966, 17.503 )
    ( 0.97, 15.146 )
    ( 0.972, 13.469 )
    ( 0.973, 12.052 )
    ( 0.974, 10.926 )
};
\addplot[line width=0.15mm,color=orange,mark=o,mark size=0.3mm]%hnsw-pca++ gist
plot coordinates {
    ( 0.92, 39.394 )
    ( 0.961, 25.853 )
    ( 0.973, 21.365 )
    ( 0.978, 17.651 )
    ( 0.981, 15.451 )
    ( 0.983, 13.044 )
    ( 0.985, 12.047 )
    ( 0.986, 10.935 )
};
\addplot[line width=0.15mm,color=forestgreen,mark=o,mark size=0.3mm]%hnsw-pca++ gist
plot coordinates {
    ( 0.924, 42.245 )
    ( 0.966, 27.655 )
    ( 0.977, 21.715 )
    ( 0.983, 18.232 )
    ( 0.986, 15.352 )
    ( 0.988, 13.318 )
    ( 0.99, 11.882 )
    ( 0.99, 10.803 )
};
\addplot[line width=0.15mm,color=pink,mark=o,mark size=0.3mm]%hnsw-pca++ gist
plot coordinates {
    ( 0.928, 41.869 )
    ( 0.969, 28.235 )
    ( 0.981, 21.973 )
    ( 0.986, 17.933 )
    ( 0.99, 15.467 )
    ( 0.992, 13.803 )
    ( 0.993, 12.255 )
    ( 0.994, 10.821 )
};
\addplot[line width=0.15mm,color=navy,mark=o,mark size=0.3mm]%hnsw-pca++ gist
plot coordinates {
    ( 0.928, 37.267 )
    ( 0.97, 26.168 )
    ( 0.982, 20.643 )
    ( 0.987, 17.068 )
    ( 0.991, 14.887 )
    ( 0.993, 13.178 )
    ( 0.994, 11.791 )
    ( 0.995, 10.682 )
};
\addplot[line width=0.15mm,color=violate,mark=o,mark size=0.3mm]%hnsw-pca++ gist
plot coordinates {
    ( 0.929, 36.161 )
    ( 0.97, 24.815 )
    ( 0.982, 20.647 )
    ( 0.988, 17.485 )
    ( 0.991, 14.867 )
    ( 0.993, 13.138 )
    ( 0.995, 11.924 )
    ( 0.995, 10.704 )
};

\end{axis}
\end{tikzpicture}\hspace{2mm}
}
\subfloat[GIST ($\HNSW$-\BSAP)]{

\begin{tikzpicture}[scale=1]
\begin{axis}[
    height=\columnwidth/2.75,
width=\columnwidth/2.0,
xlabel=recall@20,
ylabel=Qpsx10,
ylabel style={yshift=-0.5mm},
label style={font=\scriptsize},
tick label style={font=\scriptsize},
% ymajorgrids=true,
% xmajorgrids=true,
% grid style=dashed,
]
\addplot[line width=0.15mm,color=red,mark=o,mark size=0.3mm]%hnsw-pca++ gist
plot coordinates {
    ( 0.901, 51.575 )
    ( 0.943, 33.367 )
    ( 0.954, 25.534 )
    ( 0.96, 21.416 )
    ( 0.964, 18.457 )
    ( 0.966, 16.263 )
    ( 0.968, 14.682 )
    ( 0.969, 13.348 )
};
\addplot[line width=0.15mm,color=orange,mark=o,mark size=0.3mm]%hnsw-pca++ gist
plot coordinates {
    ( 0.912, 47.492 )
    ( 0.955, 30.543 )
    ( 0.967, 23.159 )
    ( 0.973, 19.397 )
    ( 0.977, 16.591 )
    ( 0.979, 14.431 )
    ( 0.981, 13.275 )
    ( 0.982, 12.068 )
};
\addplot[line width=0.15mm,color=forestgreen,mark=o,mark size=0.3mm]%hnsw-pca++ gist
plot coordinates {
    ( 0.917, 45.771 )
    ( 0.961, 28.981 )
    ( 0.972, 22.678 )
    ( 0.979, 18.617 )
    ( 0.982, 16.132 )
    ( 0.984, 14.279 )
    ( 0.986, 12.791 )
    ( 0.987, 11.944 )
};
\addplot[line width=0.15mm,color=pink,mark=o,mark size=0.3mm]%hnsw-pca++ gist
plot coordinates {
    ( 0.925, 42.434 )
    ( 0.967, 27.882 )
    ( 0.978, 21.014 )
    ( 0.985, 17.405 )
    ( 0.988, 15.023 )
    ( 0.99, 13.198 )
    ( 0.992, 11.84 )
    ( 0.993, 10.664 )
};
\addplot[line width=0.15mm,color=navy,mark=o,mark size=0.3mm]%hnsw-pca++ gist
plot coordinates {
    ( 0.926, 29.539 )
    ( 0.968, 19.357 )
    ( 0.98, 15.487 )
    ( 0.986, 13.63 )
    ( 0.99, 12.513 )
    ( 0.992, 11.255 )
    ( 0.994, 10.312 )
    ( 0.994, 9.372 )
};
\addplot[line width=0.15mm,color=violate,mark=o,mark size=0.3mm]%hnsw-pca++ gist
plot coordinates {
    ( 0.928, 30.072 )
    ( 0.969, 15.963 )
    ( 0.981, 12.699 )
    ( 0.988, 11.022 )
    ( 0.991, 10.046 )
    ( 0.993, 9.318 )
    ( 0.994, 8.998 )
    ( 0.995, 8.275 )
};

\end{axis}
\end{tikzpicture}\hspace{2mm}
}
\subfloat[DEEP ($\HNSW$-\BSAO)]{

\begin{tikzpicture}[scale=1]
\begin{axis}[
    height=\columnwidth/2.75,
width=\columnwidth/2.00,
xlabel=recall@20,
ylabel=Qpsx100,
label style={font=\scriptsize},
tick label style={font=\scriptsize},
% ymajorgrids=true,
% xmajorgrids=true,
% grid style=dashed,
]
\addplot[line width=0.15mm,color=red,mark=o,mark size=0.3mm]%hnsw-pca++ deep1M
plot coordinates {
    ( 0.919, 15.503 )
    ( 0.956, 10.186 )
    ( 0.967, 7.835 )
    ( 0.972, 6.343 )
    ( 0.974, 5.325 )
    ( 0.976, 4.583 )
    ( 0.977, 4.23 )
    ( 0.978, 4.03 )
};
\addplot[line width=0.15mm,color=orange,mark=triangle,mark size=0.3mm]%hnsw-pca++ deep1M
plot coordinates {
    ( 0.926, 15.558 )
    ( 0.965, 10.813 )
    ( 0.977, 8.191 )
    ( 0.982, 6.525 )
    ( 0.984, 5.468 )
    ( 0.986, 4.765 )
    ( 0.987, 4.32 )
    ( 0.988, 3.833 )
};
\addplot[line width=0.15mm,color=forestgreen,mark=square,mark size=0.3mm]%hnsw-pca++ deep1M
plot coordinates {
    ( 0.929, 14.036 )
    ( 0.969, 8.187 )
    ( 0.981, 6.984 )
    ( 0.986, 5.938 )
    ( 0.988, 4.816 )
    ( 0.99, 4.338 )
    ( 0.991, 3.86 )
    ( 0.992, 3.487 )
};
\addplot[line width=0.15mm,color=pink,mark=star,mark size=0.3mm]%hnsw-pca++ deep1M
plot coordinates {
    ( 0.933, 14.066 )
    ( 0.972, 9.61 )
    ( 0.985, 7.635 )
    ( 0.99, 6.447 )
    ( 0.992, 5.394 )
    ( 0.994, 4.591 )
    ( 0.995, 4.09 )
};
\addplot[line width=0.15mm,color=navy,mark=oplus,mark size=0.3mm]%hnsw-pca++ deep1M
plot coordinates {
    ( 0.933, 14.582 )
    ( 0.973, 9.722 )
    ( 0.986, 8.063 )
    ( 0.99, 6.397 )
    ( 0.993, 4.775 )
    ( 0.995, 4.336 )
};
\addplot[line width=0.15mm,color=violate,mark=otimes,mark size=0.3mm]%hnsw-pca++ deep1M
plot coordinates {
    ( 0.934, 12.933 )
    ( 0.974, 8.886 )
    ( 0.986, 7.461 )
    ( 0.991, 6.176 )
    ( 0.994, 5.218 )
    ( 0.996, 4.728 )
};

\end{axis}
\end{tikzpicture}\hspace{2mm}
}
\subfloat[DEEP ($\HNSW$-\BSAP)]{

\begin{tikzpicture}[scale=1]
\begin{axis}[
    height=\columnwidth/2.75,
width=\columnwidth/2.00,
xlabel=recall@20,
ylabel=Qpsx100,
ylabel style={yshift=-0.5mm},
label style={font=\scriptsize},
tick label style={font=\scriptsize},
% ymajorgrids=true,
% xmajorgrids=true,
% grid style=dashed,
]
\addplot[line width=0.15mm,color=red,mark=o,mark size=0.3mm]%hnsw-pca++ deep1M
plot coordinates {
    ( 0.918, 21.405 )
    ( 0.956, 12.071 )
    ( 0.966, 9.86 )
    ( 0.971, 7.8 )
    ( 0.974, 6.452 )
    ( 0.976, 5.666 )
    ( 0.976, 4.961 )
    ( 0.977, 4.467 )
};
\addplot[line width=0.15mm,color=orange,mark=traingle,mark size=0.3mm]%hnsw-pca++ deep1M
plot coordinates {
    ( 0.927, 21.277 )
    ( 0.965, 13.272 )
    ( 0.977, 9.743 )
    ( 0.981, 7.803 )
    ( 0.984, 6.511 )
    ( 0.986, 5.49 )
    ( 0.987, 5.025 )
    ( 0.988, 4.507 )
};
\addplot[line width=0.15mm,color=forestgreen,mark=square,mark size=0.3mm]%hnsw-pca++ deep1M
plot coordinates {
    ( 0.931, 21.048 )
    ( 0.969, 13.199 )
    ( 0.98, 9.621 )
    ( 0.986, 7.766 )
    ( 0.988, 6.452 )
    ( 0.99, 5.504 )
    ( 0.991, 4.939 )
    ( 0.992, 4.399 )
};
\addplot[line width=0.15mm,color=pink,mark=star,mark size=0.3mm]%hnsw-pca++ deep1M
plot coordinates {
    ( 0.932, 20.296 )
    ( 0.971, 12.129 )
    ( 0.983, 9.108 )
    ( 0.989, 7.323 )
    ( 0.991, 6.082 )
    ( 0.993, 5.267 )
    ( 0.995, 4.691 )
    ( 0.995, 4.265 )
};
\addplot[line width=0.15mm,color=navy,mark=oplus,mark size=0.3mm]%hnsw-pca++ deep1M
plot coordinates {
    ( 0.933, 20.198 )
    ( 0.972, 12.851 )
    ( 0.984, 9.474 )
    ( 0.991, 7.413 )
    ( 0.993, 6.336 )
    ( 0.995, 5.429 )
    ( 0.996, 4.811 )
};
\addplot[line width=0.15mm,color=violate,mark=otimes,mark size=0.3mm]%hnsw-pca++ deep1M
plot coordinates {
    ( 0.933, 19.862 )
    ( 0.972, 12.397 )
    ( 0.985, 9.126 )
    ( 0.992, 7.344 )
    ( 0.994, 6.136 )
    ( 0.995, 5.263 )
};

\end{axis}
\end{tikzpicture}\hspace{2mm}
}
\caption{Varying the Target Recall}\vgap\label{fig:target_recall}
\end{small}\vgap\vgap
\end{figure}

%% file: figures/result_plot_avx_small.tex
\begin{figure}[t!]
\centering
\begin{small}
\begin{tikzpicture}
    \begin{customlegend}[legend columns=3,
        legend entries={$\HNSW$,$\HNSW$++,$\HNSW$-\BSAO,$\HNSW$-\BSAP,$\HNSW$-\BSAR,FINGER},
        legend style={at={(0.45,1.15)},anchor=north,draw=none,font=\scriptsize,column sep=0.1cm}]
    \addlegendimage{line width=0.15mm,color=amaranth,mark=o,mark size=0.5mm}
    \addlegendimage{line width=0.15mm,color=amber,mark=triangle,mark size=0.5mm}
    \addlegendimage{line width=0.15mm,color=forestgreen,mark=otimes,mark size=0.5mm}
    \addlegendimage{line width=0.15mm,color=airforceblue,mark=pentagon,mark size=0.5mm}
    \addlegendimage{line width=0.15mm,color=navy,mark=oplus,mark size=0.5mm}
    \addlegendimage{line width=0.15mm,color=black,mark=halfcircle,mark size=0.5mm}
    \end{customlegend}
\end{tikzpicture}
\\[-\lineskip]

% gist@20
\subfloat[GIST (HNSW)]{

\begin{tikzpicture}[scale=1]
\begin{axis}[
    height=\columnwidth/2.70,
width=\columnwidth/1.90,
xlabel=recall@20,
ylabel=Qpsx100,
label style={font=\scriptsize},
tick label style={font=\scriptsize},
ymajorgrids=true,
xmajorgrids=true,
grid style=dashed,
]
\addplot[line width=0.15mm,color=amaranth,mark=o,mark size=0.5mm]%hnsw gist
plot coordinates {
    ( 0.934, 3.642 )
    ( 0.973, 2.348 )
    ( 0.984, 1.659 )
    ( 0.989, 1.323 )
    ( 0.992, 1.109 )
    ( 0.993, 0.959 )
    ( 0.995, 0.848 )
    ( 0.996, 0.762 )
};
\addplot[line width=0.15mm,color=amber,mark=triangle,mark size=0.5mm]%hnsw++ gist
plot coordinates {
    ( 0.931, 5.415 )
    ( 0.97, 3.417 )
    ( 0.981, 2.584 )
    ( 0.986, 2.119 )
    ( 0.989, 1.794 )
    ( 0.99, 1.593 )
    ( 0.991, 1.419 )
    ( 0.992, 1.304 )
};
\addplot[line width=0.15mm,color=forestgreen,mark=otimes,mark size=0.5mm]%hnsw-opq-sse gist
plot coordinates {
    ( 0.929, 7.423 )
    ( 0.97, 5.256 )
    ( 0.982, 3.929 )
    ( 0.988, 2.879 )
    ( 0.99, 2.736 )
    ( 0.993, 1.993 )
    ( 0.994, 2.181 )
    ( 0.995, 2.017 )
};
\addplot[line width=0.15mm,color=airforceblue,mark=pentagon,mark size=0.5mm]%hnsw-learn-pca gist
plot coordinates {
    ( 0.929, 5.739 )
    ( 0.97, 3.515 )
    ( 0.982, 2.786 )
    ( 0.988, 2.289 )
    ( 0.991, 1.964 )
    ( 0.992, 1.74 )
    ( 0.994, 1.571 )
    ( 0.995, 1.43 )
};
\addplot[line width=0.15mm,color=navy,mark=oplus,mark size=0.5mm]%hnsw-res-infer++ gist
plot coordinates {
    ( 0.929, 8.905 )
    ( 0.967, 6.028 )
    ( 0.978, 4.707 )
    ( 0.984, 3.833 )
    ( 0.987, 3.346 )
    ( 0.988, 3.002 )
    ( 0.989, 2.278 )
    ( 0.99, 1.735 )
};
%recall@20 gist
\addplot[line width=0.15mm,color=black,mark=halfcircle,mark size=0.5mm]%hnsw-finger-infer gist
plot coordinates {
    ( 0.932, 6.969 )
    ( 0.971, 3.728 )
    ( 0.979, 3.139 )
    ( 0.992, 1.561 )
};

\end{axis}
\end{tikzpicture}\hspace{2mm}
}
% gist@100
\subfloat[GIST (HNSW)]{

\begin{tikzpicture}[scale=1]
\begin{axis}[
    height=\columnwidth/2.70,
width=\columnwidth/1.90,
xlabel=recall@100,
ylabel=Qpsx100,
label style={font=\scriptsize},
tick label style={font=\scriptsize},
ymajorgrids=true,
xmajorgrids=true,
grid style=dashed,
]
\addplot[line width=0.15mm,color=amaranth,mark=o,mark size=0.5mm]%hnsw gist
plot coordinates {
    ( 0.883, 4.792 )
    ( 0.948, 2.652 )
    ( 0.97, 1.723 )
    ( 0.98, 1.118 )
    ( 0.986, 1.045 )
    ( 0.989, 0.931 )
    ( 0.991, 0.842 )
    ( 0.993, 0.757 )
};
\addplot[line width=0.15mm,color=amber,mark=triangle,mark size=0.5mm]%hnsw++ gist
plot coordinates {
    ( 0.882, 3.272 )
    ( 0.947, 2.149 )
    ( 0.969, 1.871 )
    ( 0.979, 1.612 )
    ( 0.984, 1.386 )
    ( 0.988, 1.226 )
    ( 0.99, 1.105 )
    ( 0.991, 1.012 )
};
\addplot[line width=0.15mm,color=forestgreen,mark=otimes,mark size=0.5mm]%hnsw-opq-sse gist
plot coordinates {
    ( 0.878, 6.56 )
    ( 0.945, 4.228 )
    ( 0.968, 3.589 )
    ( 0.979, 2.827 )
    ( 0.984, 2.622 )
    ( 0.988, 2.339 )
    ( 0.991, 2.084 )
    ( 0.992, 1.896 )
};
\addplot[line width=0.15mm,color=airforceblue,mark=pentagon,mark size=0.5mm]%hnsw-learn-pca gist
plot coordinates {
    ( 0.881, 4.657 )
    ( 0.945, 2.952 )
    ( 0.968, 2.231 )
    ( 0.978, 1.865 )
    ( 0.984, 1.607 )
    ( 0.987, 1.41 )
    ( 0.99, 1.279 )
    ( 0.991, 0.996 )
};
\addplot[line width=0.15mm,color=navy,mark=oplus,mark size=0.5mm]%hnsw-res-infer++ gist
plot coordinates {
    ( 0.882, 5.858 )
    ( 0.948, 3.244 )
    ( 0.969, 2.912 )
    ( 0.979, 2.579 )
    ( 0.985, 2.411 )
    ( 0.988, 2.285 )
    ( 0.99, 2.078 )
    ( 0.992, 1.914 )
};
%recall@100 gist
\addplot[line width=0.15mm,color=black,mark=halfcircle,mark size=0.5mm]%hnsw-finger-infer gist
plot coordinates {
    ( 0.873, 6.337 )
    ( 0.945, 3.653 )
    ( 0.96, 3.097 )
    ( 0.98, 2.025 )
    ( 0.986, 1.546 )
};

\end{axis}
\end{tikzpicture}\hspace{2mm}
}
\\
\vspace{1mm}
% deep1M@20
\subfloat[DEEP (HNSW)]{

\begin{tikzpicture}[scale=1]
\begin{axis}[
    height=\columnwidth/2.70,
width=\columnwidth/1.90,
xlabel=recall@20,
ylabel=Qpsx100,
label style={font=\scriptsize},
tick label style={font=\scriptsize},
ymajorgrids=true,
xmajorgrids=true,
grid style=dashed,
]
\addplot[line width=0.15mm,color=amaranth,mark=o,mark size=0.5mm]%hnsw deep1M
plot coordinates {
    ( 0.936, 25.143 )
    ( 0.975, 17.414 )
    ( 0.988, 12.234 )
    ( 0.992, 9.434 )
    ( 0.995, 7.845 )
    ( 0.996, 6.701 )
    ( 0.998, 5.874 )
    ( 0.998, 5.235 )
};
\addplot[line width=0.15mm,color=amber,mark=triangle,mark size=0.5mm]%hnsw++ deep1M
plot coordinates {
    ( 0.936, 19.154 )
    ( 0.975, 11.7 )
    ( 0.987, 8.685 )
    ( 0.992, 7.033 )
    ( 0.995, 5.956 )
    ( 0.996, 4.951 )
    ( 0.997, 4.588 )
    ( 0.998, 4.163 )
};
\addplot[line width=0.15mm,color=forestgreen,mark=otimes,mark size=0.5mm]%hnsw-opq-sse deep1M
plot coordinates {
    ( 0.933, 26.378 )
    ( 0.973, 17.631 )
    ( 0.986, 13.313 )
    ( 0.992, 11.224 )
    ( 0.994, 9.529 )
    ( 0.996, 7.821 )
    ( 0.997, 6.653 )
    ( 0.997, 6.393 )
};
\addplot[line width=0.15mm,color=airforceblue,mark=pentagon,mark size=0.5mm]%hnsw-learn-pca deep1M
plot coordinates {
    ( 0.934, 29.563 )
    ( 0.972, 18.989 )
    ( 0.985, 14.002 )
    ( 0.991, 11.391 )
    ( 0.994, 9.62 )
    ( 0.995, 8.426 )
    ( 0.996, 7.48 )
    ( 0.997, 5.943 )
};
\addplot[line width=0.15mm,color=navy,mark=oplus,mark size=0.5mm]%hnsw-res-infer++ deep1M
plot coordinates {
    ( 0.932, 36.362 )
    ( 0.972, 22.241 )
    ( 0.986, 16.445 )
    ( 0.991, 13.123 )
    ( 0.994, 10.942 )
    ( 0.995, 9.563 )
    ( 0.996, 8.476 )
    ( 0.997, 6.218 )
};
%recall@20 deep1M
\addplot[line width=0.15mm,color=black,mark=halfcircle,mark size=0.5mm]%hnsw-finger-infer deep1M
plot coordinates {
    ( 0.902, 28.007 )
    ( 0.961, 22.938 )
    ( 0.986, 12.771 )
    ( 0.996, 6.894 )
    ( 0.997, 5.67 )
    ( 0.999, 3.372 )
    ( 0.999, 2.831 )
};

\end{axis}
\end{tikzpicture}\hspace{2mm}
}
% deep1M@100
\subfloat[DEEP (HNSW)]{

\begin{tikzpicture}[scale=1]
\begin{axis}[
    height=\columnwidth/2.70,
width=\columnwidth/1.90,
xlabel=recall@100,
ylabel=Qpsx100,
label style={font=\scriptsize},
tick label style={font=\scriptsize},
ymajorgrids=true,
xmajorgrids=true,
grid style=dashed,
]
\addplot[line width=0.15mm,color=amaranth,mark=o,mark size=0.5mm]%hnsw deep1M
plot coordinates {
    ( 0.869, 30.085 )
    ( 0.944, 17.139 )
    ( 0.968, 12.202 )
    ( 0.98, 9.499 )
    ( 0.986, 7.232 )
    ( 0.989, 6.628 )
    ( 0.992, 5.808 )
    ( 0.994, 5.167 )
};
\addplot[line width=0.15mm,color=amber,mark=triangle,mark size=0.5mm]%hnsw++ deep1M
plot coordinates {
    ( 0.869, 17.212 )
    ( 0.944, 10.075 )
    ( 0.968, 7.412 )
    ( 0.98, 5.994 )
    ( 0.986, 5.064 )
    ( 0.989, 3.174 )
    ( 0.992, 2.699 )
    ( 0.994, 2.459 )
};
\addplot[line width=0.15mm,color=forestgreen,mark=otimes,mark size=0.5mm]%hnsw-opq-sse deep1M
plot coordinates {
    ( 0.869, 21.836 )
    ( 0.942, 15.165 )
    ( 0.968, 12.252 )
    ( 0.979, 10.088 )
    ( 0.985, 8.818 )
    ( 0.989, 7.579 )
    ( 0.991, 6.928 )
    ( 0.993, 6.298 )
};
\addplot[line width=0.15mm,color=airforceblue,mark=pentagon,mark size=0.5mm]%hnsw-learn-pca deep1M
plot coordinates {
    ( 0.869, 23.022 )
    ( 0.943, 14.96 )
    ( 0.967, 11.436 )
    ( 0.978, 9.459 )
    ( 0.984, 8.16 )
    ( 0.988, 7.147 )
    ( 0.99, 5.334 )
    ( 0.992, 3.958 )
};
\addplot[line width=0.15mm,color=navy,mark=oplus,mark size=0.5mm]%hnsw-res-infer++ deep1M
plot coordinates {
    ( 0.869, 28.684 )
    ( 0.943, 19.065 )
    ( 0.968, 14.354 )
    ( 0.979, 11.583 )
    ( 0.985, 9.885 )
    ( 0.989, 8.578 )
    ( 0.992, 7.569 )
    ( 0.994, 6.867 )
};
%recall@100 deep1M
\addplot[line width=0.15mm,color=black,mark=halfcircle,mark size=0.5mm]%hnsw-finger-infer deep1M
plot coordinates {
    ( 0.774, 32.014 )
    ( 0.961, 11.945 )
    ( 0.988, 6.861 )
    ( 0.992, 3.12 )
    ( 0.997, 3.776 )
    ( 0.998, 3.243 )
};

\end{axis}
\end{tikzpicture}\hspace{2mm}
}
\\
\vspace{1mm}
\caption{The Comparison with FINGER}\vgap\label{fig:avx512-time-acc-trade-small}
\end{small}\vgap
\end{figure}

%% file: figures/scan-dim-prune-rate.tex
\begin{figure}[!t]
\centering
\begin{small}
\begin{tikzpicture}
    \begin{customlegend}[legend columns=5,
        legend entries={Naive,Rand,\BSAP,\BSAO,\BSAR},
        legend style={at={(0.45,1.15)},anchor=north,draw=none,font=\scriptsize,column sep=0.1cm}]
    \addlegendimage{line width=0.15mm,color=red,mark=star,mark size=0.5mm}
    \addlegendimage{line width=0.15mm,color=orange,mark=triangle,mark size=0.5mm}
    \addlegendimage{line width=0.15mm,color=forestgreen,mark=square,mark size=0.5mm}
    \addlegendimage{line width=0.15mm,color=navy,mark=otimes,mark size=0.5mm}
    \addlegendimage{line width=0.15mm,color=violate,mark=oplus,mark size=0.5mm}
    \end{customlegend}
\end{tikzpicture}

\begin{tikzpicture}[scale=1]
\begin{axis}[
    height=\columnwidth/2.70,
    width=\columnwidth/2.0,
axis y line*=left,
xlabel=GIST-\textit{efsearch}x10,
ylabel=scan-rate\%,
label style={font=\scriptsize},
tick label style={font=\scriptsize},
every axis y label/.style={at={(current axis.north west)},right=8mm,above=0mm},
]

%Niave
\addplot[width=0.15mm,color=red,mark=star,mark size=0.5mm]
    coordinates {
    ( 25.0, 1 )
    ( 50.0, 1)
    ( 75.0, 1)
    ( 100.0, 1)
    ( 125.0, 1)
    ( 150.0, 1)
    ( 175.0, 1)
    ( 200.0, 1)
    };

%ADsampling
\addplot[width=0.15mm,color=orange,mark=triangle,mark size=0.5mm]
    coordinates {
    ( 25.0, 0.47639 )
    ( 50.0, 0.39384 )
    ( 75.0, 0.35157 )
    ( 100.0, 0.32425 )
    ( 125.0, 0.30457 )
    ( 150.0, 0.28953 )
    ( 175.0, 0.27742 )
    ( 200.0, 0.26736 )
    };
%PCA
\addplot[width=0.15mm,color=forestgreen,mark=square,mark size=0.5mm]
    coordinates {
    ( 25.0, 0.24376 )
    ( 50.0, 0.20546 )
    ( 75.0, 0.18769 )
    ( 100.0, 0.17667 )
    ( 125.0, 0.16894 )
    ( 150.0, 0.16314 )
    ( 175.0, 0.15851 )
    ( 200.0, 0.15466 )
    };
%RES
\addplot[width=0.15mm,color=violate,mark=oplus,mark size=0.5mm]
    coordinates {
    (25, 0.116733)
    (50, 0.096544)
    (75, 0.087143)
    (100, 0.081273)
    (125, 0.077119)
    (150, 0.073979)
    (175, 0.071454)
    (200, 0.069363)
    };

\end{axis}
\begin{axis}[
    height=\columnwidth/2.70,
    width=\columnwidth/2.0,
axis y line*=right,
label style={font=\scriptsize},
tick label style={font=\scriptsize},
]

\addplot[width=0.15mm,color=navy,mark=otimes,mark size=0.5mm]
    coordinates {
    ( 25.0, 0.97333 )
    ( 50.0, 0.97633 )
    ( 75.0, 0.97717 )
    ( 100.0, 0.97728 )
    ( 125.0, 0.97787 )
    ( 150.0, 0.97819 )
    ( 175.0, 0.97867 )
    ( 200.0, 0.9787 )
    };

\end{axis}
\end{tikzpicture}
\begin{tikzpicture}[scale=1]
\begin{axis}[
    height=\columnwidth/2.70,
    width=\columnwidth/2.0,
axis y line*=left,
xlabel=GIST-\textit{nprobe},
label style={font=\scriptsize},
tick label style={font=\scriptsize},
]

%Niave
\addplot[width=0.15mm,color=red,mark=star,mark size=0.5mm]
    coordinates {
    ( 50.0, 1)
    ( 100.0, 1)
    ( 150.0, 1)
    ( 200.0, 1)
    ( 250.0, 1)
    ( 300.0, 1)
    ( 350.0, 1)
    ( 400.0, 1)
    };

%ADsampling
\addplot[width=0.15mm,color=orange,mark=triangle,mark size=0.5mm]
    coordinates {
    ( 50.0, 0.24986 )
    ( 100.0, 0.19684 )
    ( 150.0, 0.17158 )
    ( 200.0, 0.15581 )
    ( 250.0, 0.14474 )
    ( 300.0, 0.13632 )
    ( 350.0, 0.12957 )
    ( 400.0, 0.124 )
    };
%PCA
\addplot[width=0.15mm,color=forestgreen,mark=square,mark size=0.5mm]
    coordinates {
    ( 50.0, 0.14559 )
    ( 100.0, 0.11876 )
    ( 150.0, 0.10656 )
    ( 200.0, 0.09911 )
    ( 250.0, 0.09388 )
    ( 300.0, 0.08992 )
    ( 350.0, 0.08674 )
    ( 400.0, 0.08409 )
    };
%RES
\addplot[width=0.15mm,color=violate,mark=oplus,mark size=0.5mm]
    coordinates {
(50, 0.068443)
(100, 0.057257)
(150, 0.052226)
(200, 0.049198)
(250, 0.047143)
(300, 0.045604)
(350, 0.044419)
(400, 0.043455)
    };

\end{axis}
\begin{axis}[
    height=\columnwidth/2.70,
    width=\columnwidth/2.0,
axis y line*=right,
ylabel=pruned-rate\%,
label style={font=\scriptsize},
tick label style={font=\scriptsize},
every axis y label/.style={at={(current axis.north east)},left=8mm,above=0mm},
ytick={},
]

\addplot[width=0.15mm,color=navy,mark=otimes,mark size=0.5mm]
    coordinates {
    ( 50.0, 0.9600461 )
    ( 100.0, 0.985783 )
    ( 150.0, 0.99269127 )
    ( 200.0, 0.9955429 )
    ( 250.0, 0.99699509 )
    ( 300.0, 0.997835 )
    ( 350.0, 0.99836493 )
    ( 400.0, 0.99872098 )
    };

\end{axis}
\end{tikzpicture}
\begin{tikzpicture}[scale=1]
\begin{axis}[
    height=\columnwidth/2.70,
    width=\columnwidth/2.0,
axis y line*=left,
xlabel=DEEP-\textit{efsearch},
ylabel=scan-rate\%,
label style={font=\scriptsize},
tick label style={font=\scriptsize},
every axis y label/.style={at={(current axis.north west)},right=8mm,above=0mm},
]
%naive
\addplot[width=0.15mm,color=red,mark=star,mark size=0.5mm]
    coordinates {
    ( 100.0, 1.0 )
    ( 200.0, 1.0 )
    ( 300.0, 1.0 )
    ( 400.0, 1.0 )
    ( 500.0, 1.0 )
    ( 600.0, 1.0 )
    ( 700.0, 1.0 )
    ( 800.0, 1.0 )
    };
%ADsampling
\addplot[width=0.15mm,color=orange,mark=triangle,mark size=0.5mm]
    coordinates {
    (100, 0.594008)
    (200, 0.523137)
    (300, 0.485096)
    (400, 0.46006)
    (500, 0.441598)
    (600, 0.427426)
    (700, 0.416202)
    (800, 0.406818)
    };
%PCA
\addplot[width=0.15mm,color=forestgreen,mark=square,mark size=0.5mm]
    coordinates {
    (100, 0.344462)
    (200, 0.319611)
    (300, 0.307854)
    (400, 0.300788)
    (500, 0.295723)
    (600, 0.291934)
    (700, 0.288931)
    (800, 0.286499)
    };
%RES
\addplot[width=0.15mm,color=violate,mark=oplus,mark size=0.5mm]
    coordinates {
    (100, 0.28429)
    (200, 0.254708)
    (300, 0.239972)
    (400, 0.230715)
    (500, 0.22381)
    (600, 0.218279)
    (700, 0.213804)
    (800, 0.210009)
    };

\end{axis}
\begin{axis}[
    height=\columnwidth/2.70,
    width=\columnwidth/2.0,
axis y line*=right,
label style={font=\scriptsize},
tick label style={font=\scriptsize},
]

\addplot[width=0.15mm,color=navy,mark=otimes,mark size=0.5mm]
    coordinates {
    ( 100.0, 0.9492 )
    ( 200.0, 0.9721 )
    ( 300.0, 0.98265 )
    ( 400.0, 0.98831 )
    ( 500.0, 0.99107 )
    ( 600.0, 0.99297 )
    ( 700.0, 0.99443 )
    ( 800.0, 0.99567 )
    };

\end{axis}
\end{tikzpicture}
\begin{tikzpicture}[scale=1]
\begin{axis}[
    height=\columnwidth/2.70,
    width=\columnwidth/2.0,
axis y line*=left,
xlabel=DEEP-\textit{nprobe},
label style={font=\scriptsize},
tick label style={font=\scriptsize},
]
%naive
\addplot[width=0.15mm,color=red,mark=star,mark size=0.5mm]
    coordinates {
    ( 50.0, 1.0 )
    ( 100.0, 1.0 )
    ( 150.0, 1.0 )
    ( 200.0, 1.0 )
    ( 250.0, 1.0 )
    ( 300.0, 1.0 )
    };
%ADsampling
\addplot[width=0.15mm,color=orange,mark=triangle,mark size=0.5mm]
    coordinates {
    (50, 0.325339)
    (100, 0.278782)
    (150, 0.256422)
    (200, 0.24212)
    (250, 0.231893)
    (300, 0.224113)
    };
%PCA
\addplot[width=0.15mm,color=forestgreen,mark=square,mark size=0.5mm]
    coordinates {
    (50, 0.246136)
    (100, 0.223237)
    (150, 0.210641)
    (200, 0.201848)
    (250, 0.195088)
    (300, 0.189612)
    };
%RES
\addplot[width=0.15mm,color=violate,mark=oplus,mark size=0.5mm]
    coordinates {
    (50, 0.188413)
    (100, 0.169941)
    (150, 0.161127)
    (200, 0.155634)
    (250, 0.151752)
    (300, 0.148808)
    };

\end{axis}
\begin{axis}[
    height=\columnwidth/2.70,
    width=\columnwidth/2.0,
axis y line*=right,
ylabel=pruned-rate\%,
label style={font=\scriptsize},
tick label style={font=\scriptsize},
every axis y label/.style={at={(current axis.north east)},left=8mm,above=0mm},
]
\addplot[width=0.15mm,color=navy,mark=otimes,mark size=0.5mm]
    coordinates {
    ( 50.0, 0.9559797 )
    ( 100.0, 0.9845504 )
    ( 150.0, 0.9920892 )
    ( 200.0, 0.99517928 )
    ( 250.0, 0.99674983 )
    ( 300.0, 0.99765796 )
    };

\end{axis}
\end{tikzpicture}
\caption{The Empirical Analysis of Our Methods}\label{fig:scan-rate-prune-rate}
\end{small}
\end{figure}

%% file: conclusion.tex
\section{Conclusion}
In this paper, we propose novel methods for distance computation in AKNN search.
We first analyze the estimation error between approximate and exact distances and prove that using PCA projection can improve the effectiveness of $\ADS$.
To further improve the generalizability of $\ADS$, we introduce a data-driven distance correction scheme that operates independently of the distance estimation source.
Extensive experiments show that our methods greatly outperform $\ADS$ in search speed.
In future work, we plan to explore advanced methods for distance computation to further refine the efficiency of AKNN search.
%\end{revise}

%\stitle{Acknowledge.}

%% file: acknowledgment.tex
\section*{Acknowledgment}
This work was supported by Ant Group Research Fund, NSFC 62302417, Guangdong Provincial Key Lab of Integrated Communication, Sensing and Computation for Ubiquitous Internet of Things (No.2023B1212010007), Guangzhou Municipal Science and Technology Project (No. 2023A03J0003, 2023A03J0013 and 2024A03J0621).